\documentclass[conference]{IEEEtran}
\usepackage{cite}
\usepackage{amsmath,amssymb,amsfonts}
\usepackage{algorithm}
\usepackage[noend,compatible]{algpseudocode}
\usepackage{graphicx}
\usepackage{textcomp}
\usepackage{xcolor}
\usepackage{fancyhdr}
\usepackage[hyphens]{url}
\usepackage{subfig}
\usepackage{soul}
\usepackage{caption}
\usepackage{comment}
\usepackage{booktabs}
\usepackage{tabularx}
\usepackage{kotex}
\usepackage{multirow}

\usepackage{xfrac}
\usepackage{makecell}
\usepackage{array}
\usepackage{siunitx}

\usepackage{hyperref}
\hypersetup{
    colorlinks=true,
    linkcolor=blue,
    filecolor=blue,
    urlcolor=blue,
    citecolor=blue,
}

\makeatletter
    \newcommand*{\algrule}[1][\algorithmicindent]{\makebox[#1][l]{\hspace*{.5em}\thealgruleextra\vrule height \thealgruleheight depth \thealgruledepth}}%
\newcommand*{\thealgruleextra}{}
\newcommand*{\thealgruleheight}{.90\baselineskip}
\newcommand*{\thealgruledepth}{.25\baselineskip}

\newcount\ALG@printindent@tempcnta
\def\ALG@printindent{%
    \ifnum \theALG@nested>0% is there anything to print
        \ifx\ALG@text\ALG@x@notext% is this an end group without any text?
        \else
            \unskip
            \addvspace{-1pt}% FUDGE to make the rules line up
            \ALG@printindent@tempcnta=1
            \loop
                \algrule[\csname ALG@ind@\the\ALG@printindent@tempcnta\endcsname]%
                \advance \ALG@printindent@tempcnta 1
            \ifnum \ALG@printindent@tempcnta<\numexpr\theALG@nested+1\relax% can't do <=, so add one to RHS and use < instead
            \repeat
        \fi
    \fi
    }%
\usepackage{etoolbox}
\patchcmd{\ALG@doentity}{\noindent\hskip\ALG@tlm}{\ALG@printindent}{}{\errmessage{failed to patch}}
\makeatother

\newbox\statebox
\newcommand{\myState}[1]{%
    \setbox\statebox=\vbox{#1}%
    \edef\thealgruleheight{\dimexpr \the\ht\statebox+1pt\relax}%
    \edef\thealgruledepth{\dimexpr \the\dp\statebox+1pt\relax}%
    \ifdim\thealgruleheight<.75\baselineskip
        \def\thealgruleheight{\dimexpr .75\baselineskip+1pt\relax}%
    \fi
    \ifdim\thealgruledepth<.25\baselineskip
        \def\thealgruledepth{\dimexpr .25\baselineskip+1pt\relax}%
    \fi
    \State #1%
    \def\thealgruleheight{\dimexpr .75\baselineskip+1pt\relax}%
    \def\thealgruledepth{\dimexpr .25\baselineskip+1pt\relax}%
}
\def\BibTeX{{\rm B\kern-.05em{\sc i\kern-.025em b}\kern-.08em
    T\kern-.1667em\lower.7ex\hbox{E}\kern-.125emX}}

\newcommand{\pluseq}{\mathrel{+}=}

\title{HEAAN Demystified: Accelerating Fully Homomorphic Encryption Through Architecture-centric Analysis and Optimization}
\author{
  \IEEEauthorblockN{Wonkyung~Jung\IEEEauthorrefmark{1}, Eojin~Lee\IEEEauthorrefmark{1}, 
  Sangpyo~Kim\IEEEauthorrefmark{1}, Keewoo~Lee\IEEEauthorrefmark{1},
  Namhoon~Kim\IEEEauthorrefmark{1},
  Chohong~Min\IEEEauthorrefmark{2}, \\
  Jung~Hee~Cheon\IEEEauthorrefmark{1}, and Jung~Ho~Ahn\IEEEauthorrefmark{1}}
  \IEEEauthorblockA{
    \IEEEauthorrefmark{1}Seoul National University,
    \IEEEauthorrefmark{2} Ewha Woman's University\\
    gajh@snu.ac.kr
  }
}

\begin{document}
\maketitle
\pagestyle{plain}

%%%%%% -- PAPER CONTENT STARTS-- %%%%%%%%

\begin{abstract}

Homomorphic Encryption (HE) draws a significant attention as a 
privacy-preserving way for cloud computing because it allows computation
on encrypted messages called ciphertexts.
Among numerous HE schemes proposed, HE for Arithmetic of Approximate Numbers 
(HEAAN) is rapidly gaining popularity across a wide range of applications
(e.g., machine learning) because it supports messages that can tolerate 
approximate computation with no limit on the number of arithmetic operations
applicable to the corresponding ciphertexts.

A critical shortcoming of HE is the high computation complexity of ciphertext
arithmetic; especially, HE multiplication (HE Mul) is more than 10,000 times slower
than the corresponding multiplication between unencrypted messages.
This leads to a large body of HE acceleration studies, including ones exploiting
FPGAs; however, those did not conduct a rigorous analysis of computational
complexity and data access patterns of HE Mul.
Moreover, the proposals mostly focused on designs with small parameter sizes,
making it difficult to accurately estimate the performance of the HE accelerators
in conducting a series of complex arithmetic operations.

In this paper, we first describe how HE Mul of HEAAN is performed in a 
manner friendly to computer architects.
Then we conduct a disciplined analysis on its computational and memory-access 
characteristics, through which we (1) extract parallelism in the key functions
composing HE Mul and
(2) demonstrate how to effectively map the parallelism to the popular parallel
processing platforms, multicore CPUs and GPUs, by applying a series of
optimization techniques such as transposing matrices and pinning data to threads.
This leads to the performance improvement of HE Mul on a CPU and a GPU
by 42.9$\times$ and 134.1$\times$, respectively, over the single-thread reference 
HEAAN running on a CPU.
The conducted analysis and optimization would set a new foundation for future HE
acceleration research.
\end{abstract}

\section{Introduction}
\label{sec:introduction}

As cloud computing becomes an increasingly dominant way of providing
computing resources, numerous computations are performed on datacenter
servers rather than on personal devices~\cite{comm-2008-cloud,comm-2010-view}.
It enables the client without expensive hardware to receive services
that require complex computations.
However, security and privacy issues are also emerging with the growth of cloud 
computing~\cite{aina-2010-cloud,nca-2011-cloud}.
When a client sends private data to a server, security issues 
in data transfers can be resolved by sending the data after encryption.
However, the data encoded by a conventional encryption method
must be decrypted to perform the computation in the server.
Therefore, a user have no choice but to use the cloud service with a 
risk of security or privacy attack (e.g., abusing) that occurs during 
the computation of unencrypted data.

Homomorphic Encryption (HE)~\cite{fsc-1978-he}, an encryption scheme that 
enables computation between encrypted data, draws significant attention as
a solution to this privacy problem.
By adopting HE, service providers no longer need to decrypt the clients'
private data for computation.
The concept of HE was first suggested in 1978~\cite{fsc-1978-he}.
However, the early proposals of HE were either unsafe~\cite{fsc-1978-he} or support only one type of HE operation,
namely HE addition (HE Add) or HE multiplication (HE Mul) (e.g., ElGamal~\cite{ieee-1985-elgamal} and
Paillier~\cite{ictact-1999-pascal}).
In this aspect, it was difficult to put HE into serious applications for a while.
However, fully HE (FHE)~\cite{stanford-2009-gentry} proposed 
in 2009 made a major breakthrough by supporting both HE Add and HE Mul.
Moreover, FHE supports bootstrapping, a method of initializing noise in
encrypted data, enabling an unbounded number of HE Add and Mul without
decryption.

Among numerous FHE schemes to date~\cite{acm-toct-2014-bgv, acm-stc-2012-fly,
iacr-2012-bfv, ictacis-2017-heaan, aictact-2015-fhew, jc-2018-tfhe}, HE for Arithmetic of 
Approximate Numbers (HEAAN~\cite{ictacis-2017-heaan}), also known as CKKS 
(Cheon-Kim-Kim-Song) is rapidly gaining popularity~\cite{idash-2018} as
it supports the approximate computation of real numbers.
HEAAN enables HE Add and Mul of approximate data, %(e.g., floating-point numbers),
where the result is almost the same as that of the original
operation with a tiny error.
Using HEAAN, computations % (e.g., deep learning) 
can be performed without data decryption in the datacenter.
However, the execution time for computation on encrypted data (ciphertext)
increases by from 100s to 10,000s of times compared
to that on native, unencrypted messages.
Therefore, it is highly desired to reduce the computation time of HE operations
to use HE practically.

There have been a large body of research on accelerating HE operations using FPGAs
or GPUs.
However, FPGA-based acceleration studies targeted the HE schemes (e.g., 
BGV~\cite{acm-toct-2014-bgv}, LTV~\cite{acm-stc-2012-fly}, and BFV~\cite{iacr-2012-bfv}) 
that only support computations of integer 
numbers~\cite{iwches-2015-ltv,ieee-2016-customaccelerator,ieee-2016-coprocessor,ieee-2018-hepcloud,ieee-2019-fpga}, 
or operate with only limited parameter sizes~\cite{arxiv-2019-heax}.
They all target performing a small number of HE Mul without bootstrapping,
inhibiting their applicability to a wide range of applications requiring
hundreds to thousands of multiplication be performed (e.g., deep learning).
GPU implementation 
studies~\cite{iccis-2015-cuhe,gpgpu-2016-yashe,iacr-2018-fvgpu,ieee-2019-rnsvariants}
do not take advantage of the algorithm's internal parallelism sufficiently,
operate on only small or limited parameters, or do not consider the cost of modulo
operations.

In this paper, we demystify HEAAN, a representative FHE scheme,
by describing, analyzing, and optimizing it in a manner friendly to
computer architects.
We first explain the pertinent details of the encryption, decryption, and
computation aspects of HEAAN, and identify that the following four
functions take more than 95\% of HE Mul, the most computationally 
expensive operation of HE: CRT (Chinese Remainder Theorem), 
NTT (Number Theoretic Transform), iNTT (inverse NTT), and iCRT (inverse CRT).
%
% To accelerate HE Mul,
we conduct an in-depth and disciplined analysis of the aforementioned
primary functions to understand their computational complexity and access
patterns on input, output, and precomputed data, which are critical for
operation (e.g., modular multiplication) strength reduction, across a 
range of key HE parameters.

The parallelism exposed through the analysis is exploited to accelerate
HE Mul on CPU and GPU, the most popular computing platforms, which are
already equipped with hundreds to thousands of ALUs.
In CPU, we utilize multiple cores (inter-core 
parallelism) and AVX-512 instructions supported by the latest Intel
architectures (intra-core parallelism).
In GPU, we utilize massive thread-level parallelism expressible through
the CUDA programming model.
We further improve % CPU and GPU
performance by proposing a 
series of architecture-centric optimizations, such as matrix 
transposition to better exploit memory access locality, loop 
reordering to expose more parallelism,
and taking a synergy between precomputation and delayed modulo
operations; and we
estimate how much more performance gains are attainable through
natively support currently emulated instructions.
We achieve 42.9$\times$ and 134.1$\times$ speedup of HE Mul on CPU 
and GPU, respectively, compared to the single-thread reference 
HEAAN~\cite{repo-2019-heaan}, setting a new baseline for the future HE
acceleration research.

\section{Background: Computational Challenges of Homomorphic Encryption (HE)}
\label{sec:background}

HE can be categorized into two groups, somewhat HE (SHE)
and fully HE (FHE), by whether there is a limitation on
the number of arithmetic operations applicable to the ciphertext.
In a HE scheme, noise is accumulated during each operation; this makes the
ciphertext of a SHE scheme indecipherable after performing a certain number
of operations.
On the contrary, FHE schemes support a bootstrapping 
algorithm~\cite{aictact-2018-bootstrapping}, which refreshes the accumulated noise.
Therefore, although there is an upper bound in the number of arithmetic 
operations that can be consecutively applied to a ciphertext, by periodically 
bootstrapping it, we can continue manipulating the ciphertext with no need for
decrypting it.
This property makes FHE well-tailored to meet the demands of a wide range of 
general applications (e.g., training/inference of deep neural 
networks~\cite{icml-2016-cryptonet, arxiv-2018-fastercryptonet, arxiv-2018-hcnn, wahc-2019-ngraph}), 
which require a massive number of operations applied to encrypted data.

Representative FHE schemes include 
Brakerski-Gentry-Vaikuntanathan (BGV)~\cite{acm-toct-2014-bgv}, 
Lopez-Alt, Tromer, and Vaikuntanathan (LTV)~\cite{acm-stc-2012-fly}, 
Brakerski/Fan-Vercauteren (BFV)~\cite{iacr-2012-bfv},
fast FHE over the torus (TFHE)~\cite{jc-2018-tfhe},
and Cheon-Kim-Kim-Song (CKKS)~\cite{ictacis-2017-heaan}. 
Among these, only CKKS supports approximate computation
on real numbers, and hence it is a top candidate for many
real-world applications
requiring a bunch of operations on data that can
tolerate tiny errors due to approximate computation.
CKKS is rapidly gaining popularity in a wide range of
applications exploiting HE, 
such as machine learning~\cite{bmcmg-2018-logistic}. 
For example, the winner and the most runner-ups of a recent HE
challenge about secure genome analysis competition (iDASH
2018~\cite{idash-2018} and 2019~\cite{idash-2019}) used CKKS
or its hybrid versions. % with other schemes.
Therefore, %in this paper,
we investigate HEAAN (HE for Arithmetic of
Approximate Numbers) scheme~\cite{repo-2019-heaan}, which is developed
by the authors of CKKS.

HEAAN is able to perform arbitrary computation types by combining HE multiplication 
(simply \emph{HE Mul}) and HE addition (simply \emph{HE Add}), which are
multiplication and addition on ciphertexts.
However, the execution time of HE operations increases significantly compared to the 
corresponding ones on the original unencrypted messages.
Table~\ref{table:mvsc} compares the execution time for addition/multiplication of 
original messages and ciphertexts using a single core from the system specified in 
Section~\ref{sec:experimental-setup}.
We measure the average execution time of addition/multiplication of a complex 
number in a message consisting of 32,768 complex numbers.
HE Mul (Add) is 36,112$\times$ (168$\times$) slower than native message 
operation.
When the message consists of fewer numbers, the slowdown is even higher.
Considering that most approximate number operations consist of multiplication and addition, 
the long execution time of HE operations is an obstacle to the practical use of HE.
Therefore,
it is essential to accelerate HE operations, 
especially HE Mul as it is 448$\times$ slower than HE Add.

\begin{table}[tb!]
  \centering
  \caption{Execution time for addition and multiplication of 
  messages vs. ciphertexts in HEAAN.
  }
  \small
  \begin{tabular}{lrrr}
    \toprule
    \textbf{Operation type} & \textbf{Message} & \textbf{Ciphertext} & \textbf{Slowdown}\\
    \midrule
    Addition & 2.1 ns& 348.2 ns & 168.2$\times$\\
    Multiplication & 4.3 ns & 155883.8 ns& 36112.7$\times$\\
    \bottomrule
  \end{tabular}
  \label{table:mvsc}
\end{table}

\section{A Brief Introduction to HEAAN}
\label{sec:introduction-to-heaan}

Prior to accelerating the most compute-intensive HEAAN operation,
HE Mul, we introduce the pertinent details of HEAAN, focusing on
how to convert an input message to a ciphertext through encoding/encryption
steps and how to perform arithmetic operations on ciphertext in HEAAN.

\subsection{HEAAN encryption}
\label{sec:heaan-encryption}

HEAAN converts an input message to a ciphertext through encoding and encryption steps.
An input message consists of $n$ complex numbers, each %where each complex number is
composed of a double-type real and imaginary number.
The encoding step first converts an input message to a plaintext 
($\textbf{t}$), a polynomial of at most degree ($N$-1) with $N$ integer coefficients. 
$\textbf{t}$ is placed in a cyclotomic polynomial ring 
($R=\mathbb{Z} [X]/(X^N + 1)=c_0X^0 + c_1X^1 + ... + c_{(N-1)}X^{(N-1)}$)
space with the magnitude of each coefficient 
bounded by ciphertext modulus, integer $q$ ($\textbf{t} \in R/qR$).
Therefore, each coefficient ($c_k$) is the residue number by $q$,
where $q$ is a BigInt (big integer) much larger than $2^{64}$.
The encoding step converts the floating-point numbers of an message %original message
to integer numbers after multiplying with a scaling factor
($\Delta$) and then rounding down the remaining fraction numbers.
Each coefficient is a log$q$-bit BigInt, % (log$q$ $\gg$ 64),
and $n \leq \frac{N}{2}$.

Then, the encryption step converts a plaintext to a ciphertext consisting of
a pair of polynomials $\textbf{c.ax}$ and $\textbf{c.bx}$ using a public key
pair ($\textbf{pk}_0$ and $\textbf{pk}_1$) as follows:
\begin{itemize}
  \item[] $\textbf{c.ax} = \textbf{u} \times \textbf{pk}_1 + \textbf{e}_1$
  \item[] $\textbf{c.bx} = \textbf{u} \times \textbf{pk}_0 + \textbf{e}_0 + \textbf{t}$
\end{itemize}
The public key pair is generated from a secret key ($\textbf{sk}$).

In the equations above,
$\textbf{u}$ is a polynomial of at most degree ($N$-1), where each coefficient is either
-1, 0, or 1 following a distribution described in~\cite{ictacis-2017-heaan}.
$\textbf{e}_0$ and $\textbf{e}_1$ are also polynomials of at most degree ($N$-1) polynomials with random error
values to ensure security, which follow a Gaussian distribution with a small standard
deviation value (e.g., $\sigma$ = $3.2$ in~\cite{ictacis-2017-heaan}).

To extract the original message from a ciphertext, we first convert the
ciphertext to the plaintext exploiting the following relationship between
$\textbf{c.ax}$ and $\textbf{c.bx}$:
\begin{itemize}
  \item[] $\textbf{c.bx} = \textbf{c.ax} \times \textbf{sk} + \textbf{t} + \textbf{e'}$
\end{itemize}
Then, the plaintext $\textbf{t}$ can be returned to the original message
through decoding; in this case, the inverse of scaling factor ($1/\Delta$)
is multiplied to get the approximate values.

HEAAN limits the maximum size of the ciphertext modulus $q$ to a constant
value $Q$. %, which determines how many HE Mul operations can be 
% performed to a ciphertext.
%
HEAAN chooses $p^L$ for $Q$, where
$L$ is multiplicative depth, the number of consecutive
HE Mul operations applicable to a ciphertext before it
loses encrypted data, and $p$ is rescaling factor.
The size of the message contained in a ciphertext increases exponentially
as the ciphertext is multiplied repeatedly.
To prevent the explosion of message size, HEAAN performs rescaling after
each HE Mul by dividing the coefficients of the output ciphertext
by $p$.
Then the size of $q$, the ciphertext modulus, is adjusted to
$q'$ where log$q'$ = log$q$ - log$p$.
Therefore, log$q$ of a ciphertext that is just encrypted starts at log$Q$,
decreases by log$p$ every time an HE Mul is applied,
and becomes 0 after experiencing $L$ HE Mul operations, losing
data.
When $p$ is fixed, more HE Mul can be applied to a
ciphertext with a larger $Q$ value.

\begin{table}[tb!]
  \centering
  \caption{Multiplicative depth ($L$) and required $N$
  over log$Q$ %values
  to guarantee 80-bit security level (an
  attacker needs $2^{80}\times$ of modular mul \& add to break with the current best algorithm.)
  }
  \small
  \begin{tabular}{rcc}
    \toprule
    log$Q$  & Multiplicative depth ($L$) & required $N$ \\
    \midrule
    300     & 10  & $2^{14}$ \\
    600     & 20  & $2^{15}$ \\
    1,200    & 40  & $2^{16}$ \\
    2,400    & 80  & $2^{17}$ \\
    \bottomrule
  \end{tabular}
  \label{tab:Q-N}
  \vspace{-0.12in}
\end{table}

To apply more HE Mul operations to a ciphertext, FHE re-initialize
the ciphertext through bootstrapping.
However, bootstrapping is a costly operation (reported to take in the order of
minutes~\cite{iacr-2018-faster}) because it consists of dozens of HE 
Mul\footnote{This again reinforces the importance of accelerating HE
Mul.} and shift operations.
To reduce the overhead of bootstrapping for practical use of HE, we should
use large $Q$ values to increase $L$.
Also, $N$ should increase as $Q$ increases to guarantee a certain 
level of security in HE (see Table~\ref{tab:Q-N}).
As larger $Q$ and $N$ values require more computation and data storage
costs per HE Mul, it is not efficient to use too large $Q$;
we discuss a further detail of this trade off in Section~\ref{sec:discussion}.
In this paper, we use ($p$, $L$, $Q$, $N$) of ($2^{30}$, 40, $2^{1200}$, $2^{16}$), 
respectively, which are the default values the official HEAAN 
repository~\cite{repo-2019-heaan} uses. % (see Table~\ref{table:formatting}).

\subsection{HEAAN computation}
\label{sec:heaan-computation}

Arithmetic operations in HEAAN, HE Add and Mul, are
through computation between the polynomials of operand ciphertexts.
Here we assume that the two operand ciphertexts in HE operation
have the same ciphertext modulus value $q$.

HE Add computes an output
ciphertext ($\textbf{c3}$) from two input ciphertexts ($\textbf{c1}$ and 
$\textbf{c2}$) through the following operations:
\begin{itemize}
  \item[] $\textbf{c3.ax} = \bmod(\textbf{c1.ax} + \textbf{c2.ax}, q)$
  \item[] $\textbf{c3.bx} = \bmod(\textbf{c1.bx} + \textbf{c2.bx}, q)$
\end{itemize}
HE Add is relatively simple because it performs the element-wise
addition of BigInt coefficients and then the modulo of $q$ for each
output coefficient; $\bmod(x, y)$ means $x$ modulo $y$.
Here the modulo operation is implemented simply by subtracting $q$ when
each output coefficient is larger than $q$ because the result of each
addition is smaller than $2q$.

\begin{figure}[!tb]
  \center
  \includegraphics[width=0.95\columnwidth]{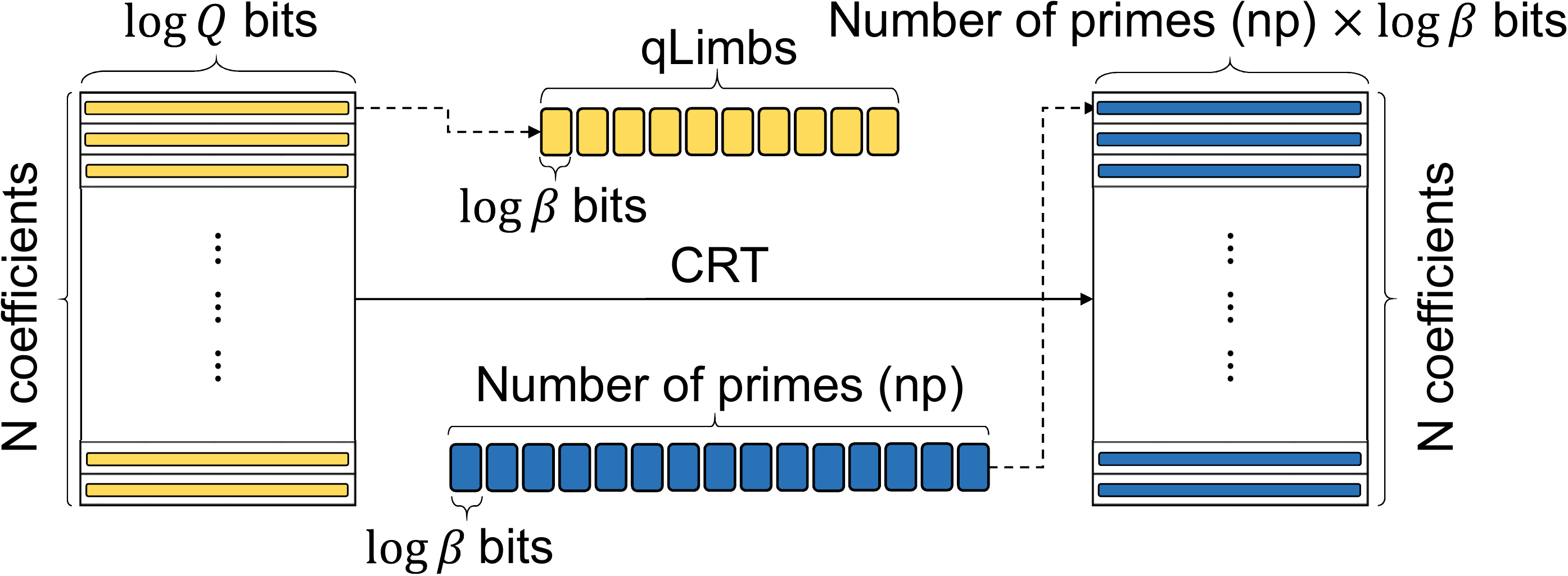}
  \vspace{-0.05in}
  \caption{
    Input and output data structure of CRT.
  }
  \label{fig:crt}
  \vspace{-0.12in}
\end{figure}

HE Mul is much more costly than HE Add because the former
requires multiplying BigInt coefficients by $N^2$ times.
In this paper, we assume that $\beta$ is the size of an integer data type
that a computer natively supports with high performance (e.g.,
$\beta$=$2^{64}$ for 64-bit CPUs), which is often called a word size.
A log$q$-bit BigInt is represented as $qLimbs$ 
(=$\lceil$log$q/$log$\beta\rceil$) words (see Figure~\ref{fig:crt}).
Then, one BigInt multiplication (simply mul) consists of $(qLimbs)^2$ log$\beta$-bit
word mul and $2(qLimbs)^2\!-\!1$ word addition (simply add) operations.
For example, because $qLimbs$ is 19 when using the representative 
parameters ($N$=$2^{16}$, log$q$=1,200, log$\beta$=64) summarized in Table~\ref{table:formatting},
361 64-bit mul and 721 64-bit add operations are required
besides carry propagation per BigInt mul.
As described above, polynomial mul requires $N^2$ (=4.3 billion)
BigInt mul, which requires at least 4.6 Tera 64-bit operations.
To reduce the complexity of this polynomial mul, HEAAN and other
HE schemes use \emph{Chinese Remainder Theorem} (CRT~\cite{world-1996-crt})
and \emph{Number Theoretic Transform} (NTT~\cite{mc-1965-ntt}).

\begin{table*}[tb!]
  \centering
  \renewcommand{\arraystretch}{1.2}
  \caption{Representative parameters in Homomorphic Encryption (HE)}
  \label{table:formatting}
  \begin{tabular}{cp{4.9in}l}
    \toprule
    \textbf{Symbol} & \textbf{Description} & \textbf{Representative values}\\
    \midrule
    $\Delta$ & \textbf{Scaling factor} is multiplied to the floating-point number of message to convert to integer number. & $\log{\Delta} = 30$\\
    $p$ & \textbf{Rescaling factor} linearly reduces the size of messages that grow exponentially during computation. & $\log{p} = 30$\\
    $L$ & \textbf{Multiplicative depth} is the maximum number of possible mul of a ciphertext without bootstrapping. & 40\\
    ${Q}$ & \textbf{The maximum ciphertext modulus} is equal to the initial ciphertext modulus after encryption.& $\log{Q} = $1,200\\
    ${q}$ & \textbf{Ciphertext modulus} starts from $Q$ and divided by a rescaling factor $p$ at each mul. & $\log{q} = $ 1,200, 1,170,$\dots$, 0\\
    ${N}$ & \textbf{The number of coefficients of ciphertext polynomial}. The degree of ciphertext polynomial is at most $N$-1. & $2^{16} = $ 65,536\\
    ${n}$ & \textbf{The number of messages}. $n$ messages are encrypted in one ciphertext.& 32, 64,$\dots$, 32,768\\ %$n$ should be a power of 2, and an operation on single ciphertext is equivalent to operation on $n$ messages in parallel.
    ${\beta}$ & \textbf{Word size}. It is machine-dependent ($2^{64}$ for CPU, $2^{32}$ for GPU). & $2^{64}$ or $2^{32}$\\
    $qLimbs$ & \textbf{The number of limbs of $q$}. To represent $\log{q}$-bit integer, $qLimbs$ limbs are required. & $\lceil$1,200/64$\rceil =$ 19\\ %\lceil \log{q}/\log{\beta} \rceil$\\
    ${np}$ & \textbf{The number of prime numbers}. We use $np$ prime numbers to represent big integer in RNS domain. & $\lceil$2,400/58$\rceil =$ 42\\%Depending on the size of big integer and 32/64-bit data type\\
    $P$ & \textbf{Product of prime numbers} that are used to represent big integer. & $P = \prod_{j}p_j$\\
    $PLimbs$ & \textbf{The number of limbs of $P$}. To represent $\log({P/p_j})$-bit integer, $PLimbs$ limbs are used.&
    $\max\limits_{j}(\log({P/p_j})/\log{\beta})$\\
    \bottomrule
  \end{tabular}
  \vspace{-0.12in}
\end{table*}

\noindent
\textbf{CRT for reducing the complexity of BigInt mul:}
CRT states that for $np$ integers $\{m_i | 0 \! \leq \! i \! < \! np\}$ which are
coprime with each other, the residue set $\{x_i$ = $\bmod(X, m_i) | 0 \! \leq \! i \! < \! np\}$ of
any integer $0 \! \leq \! X \! < \! \prod_{i=0}^{np-1}m_i$ is unique.
HEAAN exploits CRT by defining a set of $np$ integers $\{p_0, p_1,...,p_{np-1}\}$, 
where each modulus $p_i$ is a prime number smaller than
$\beta$ and $\prod_{i=0}^{np-1}p_i$ = $P \! \geq \! q^2$.
Then, a log$q$-bit BigInt number $B$, which is the coefficient of the ciphertext 
polynomial, can be represented in the residue number
system (RNS) by the set of remainders $\{b_0, b_1,...,b_{np-1}\}$ where
$b_i$ =$\bmod(B, p_i)$.

A key property of RNS is that for adding, subtracting, and multiplying numbers
represented in RNS, it is sufficient to perform the same modular operation on
each pair of residues (called a \emph{congruence relation}).
That is, for a pair of log$q$-bit BigInt numbers ($A$, $B$) and their
corresponding RNS representations 
($\{a_i | 0 \!\leq\!i\!<np\}$, $\{b_i | 0 \!\leq\!i\!<np\}$), the product of
$A$ and $B$ is $C$ represented by $\{c_i | 0 \!\leq\!i\!<np\}$ such that
$c_i$ =$\bmod(a_i \cdot b_i, p_i)$.
This relationship holds because we set $P$ not smaller than $q^2$ and the
product of two log$q$-bit BigInt numbers are smaller than $q^2$.

Therefore, a log$q$-bit BigInt number is converted into $np$ log$\beta$-bit
data (see Figure~\ref{fig:crt}) in an RNS domain, where we call the
conversion \emph{a CRT function} or simply \texttt{CRT}, and a BigInt 
mul is changed into $np$ log$\beta$-bit modular mul;
and hence, the time complexity per BigInt mul is changed from
$\mathcal{O}(qLimbs^2)$ to $\mathcal{O}(np)$.
In general, $qLimbs^2 \gg np$ (see Table~\ref{table:formatting}), so the
number of operations required for BigInt mul can be greatly
reduced by \texttt{CRT}.
However, multiplying two BigInt polynomials still has a
complexity of $\mathcal{O}(np \cdot N^2)$ because polynomial mul requires
$N^2$ coefficient mul operations.

\noindent
\textbf{NTT for reducing the complexity of polynomial mul:}
NTT is a discrete Fourier transform over a finite field (integer).
It is well known that fast polynomial mul can be implemented 
by using Fast Fourier Transform (FFT)~\cite{iants-2008-fftpoly}. 
Therefore, we can translate polynomial mul with $\mathcal{O}(N^2)$ complexity 
into element-wise mul with $\mathcal{O}(N)$ complexity by using fast NTT, a
variant of FFT limited to integer values. 
Although fast NTT (or simply \texttt{NTT}) requires transformation cost with
$\mathcal{O}(N\log{N}$) complexity, it is beneficial to use \texttt{NTT} when $N$ is
large enough.

\begin{figure}[!tb]
  \center
  \includegraphics[width=\columnwidth]{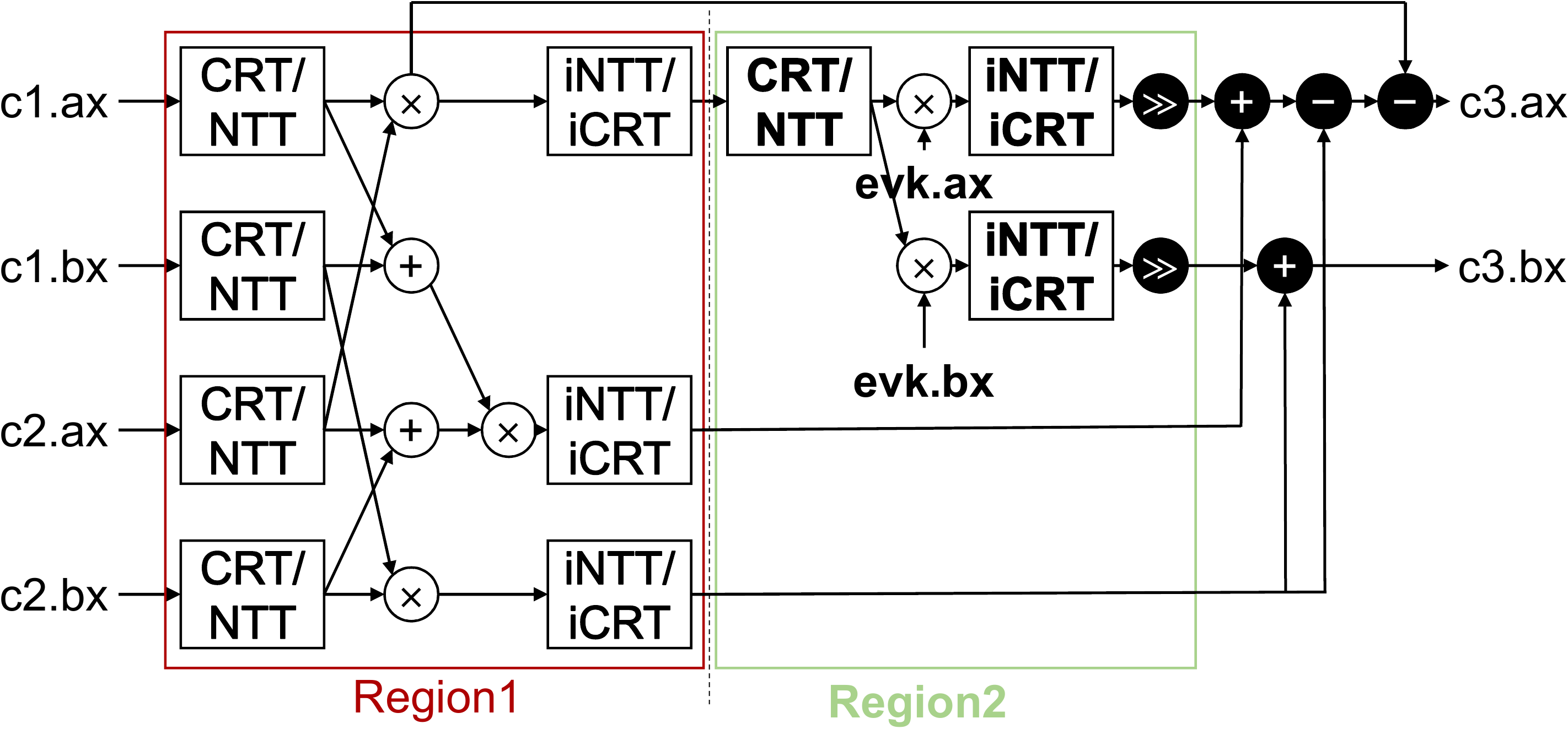}
  \vspace{-0.15in}
  \caption{
    The overall flow of HE Mul in HEAAN.
    A white (black) filled symbol represents an operation conducted in an RNS (BigInt) domain.
    Region 1 and region 2 use different moduli because the former 
    multiplies two log$q$-bit numbers whereas the latter multiplies
    a log$q$-bit number with an evaluation key polynomial composed of
    log$Q^2$-bit numbers.
  }
  \label{fig:he_mult_flow}
  \vspace{-0.07in}
\end{figure}

The overall flow of HE Mul in HEAAN, including \texttt{CRT} and 
\texttt{NTT}, is depicted in Figure~\ref{fig:he_mult_flow}.
It consists of 5 polynomial mul, where each polynomial 
mul performs (1) \texttt{CRT}, (2) \texttt{NTT}, and (3)
element-wise modular mul, followed by (4) inverse fast
NTT (\texttt{iNTT}) and (5) RNS-to-BigInt conversion (\texttt{iCRT})
to return to the polynomial with BigInt coefficients.
We divide the whole process into region 1 and region 2; the former
multiplies and adds input ciphertexts whereas the latter transforms $\textbf{c1.ax} \times \textbf{c2.ax}$,
which is a region 1's byproduct that is decrypted with $\textbf{sk}^2$, 
into a form that is decrypted with $\textbf{sk}$ (called $\textit{key-switching}$).
The evaluation key ($\textbf{evk}$) used in region 2 consists of two
polynomials ($\textbf{evk.ax}$ and $\textbf{evk.bx}$) where
each coefficient is log$Q^2$-bit (= 2$\cdot$log$Q$-bit) long,
encrypting the square of $\textbf{sk}$ multiplied by a constant Q.
The key-switching procedure with $\textbf{evk}$ is mandatory to decrypt correctly,
as it removes the need for $\textbf{sk}^n$ in decryption stage,
which, without the key-switching, would be required after $n$ sequential multiplications.

In region 1, $np$ is configured to deal with log$q^2$-bit %(= 2$\times$log$q$-bit)
BigInt, the intermediate result of polynomial mul between
two input ciphertexts whose coefficient size is log$q$-bit.
By contrast, in region 2, $np$ is set larger to represent (log$q$+
log$Q^2$)-bit BigInt because the coefficient size of
$\textbf{evk}$ is log$Q^2$-bit long.
The shift operations in region 2 reduce the amount of error
that was accumulated during mul. 
The following additions and subtractions between the results of 
polynomial mul produce the result of HE Mul 
($\textbf{c3.ax}$ and $\textbf{c3.bx}$); we can get an approximate 
value of mul between two original messages by
decrypting the result using $\textbf{sk}$.

\begin{figure}[!tb]
  \center
  \includegraphics[width=\columnwidth]{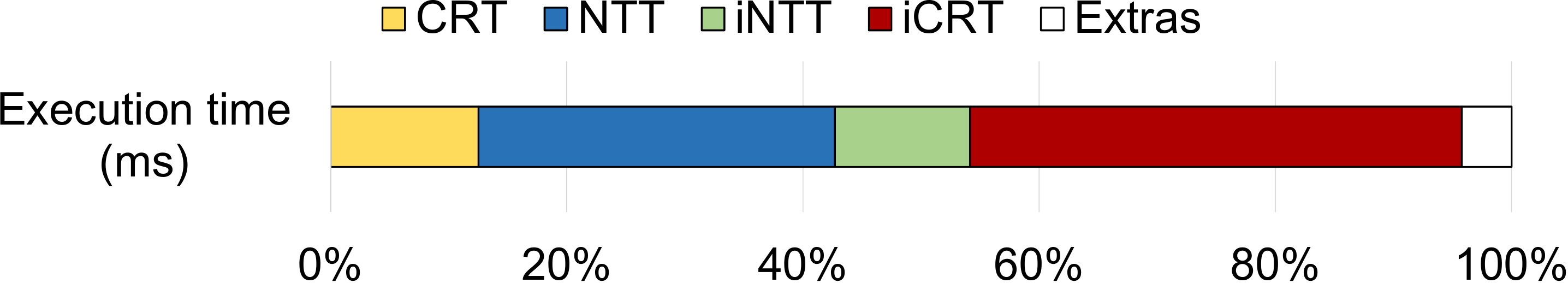}
  \vspace{-0.15in}
  \caption{
    HE Mul execution time breakdown (total 5,108 ms).
  }
  \label{fig:he_mult_time}
  % \vspace{-0.12in}
\end{figure}

Figure~\ref{fig:he_mult_time} shows the execution time breakdown of HE 
Mul using a single-threaded reference HEAAN in the system and configuration 
described in Section~\ref{sec:experimental-setup}.
\texttt{CRT}, \texttt{NTT}, \texttt{iNTT}, and \texttt{iCRT} account for
95.8\% of the total execution time.
The remaining operations, such as element-wise modular mul, 
account for only 4.2\% of execution time.
The total execution time is 5,108 ms, which is about 36,000$\times$ slower than 
the original message mul, as discussed in Section~\ref{sec:background}.
\emph{Therefore, accelerating HE Mul is essential for practical use of 
HE, and it is necessary to accelerate \texttt{CRT}, \texttt{NTT}, 
\texttt{iNTT}, and \texttt{iCRT}.}

\section{An In-depth Analysis of Major Functions in HEAAN Multiplication}
\label{sec:contribution-2}

To accelerate the primary functions (\texttt{CRT}, \texttt{NTT}, \texttt{iNTT},
and \texttt{iCRT}) in HE Mul, we first conduct an in-depth analysis
of how each function works.
In the following descriptions, $\mathbf{IN}/\mathbf{OUT_{function}} (X, Y)$ 
represents an $X$ by $Y$ matrix used as input/output of a function, while 
$\mathbf{TB_{function}} (X, Y)$ represents a precomputed table 
of $X$ by $Y$ matrix.

\begin{algorithm}[tb!]
\caption{CRT}
\small
\label{algorithm:crt}
\begin{algorithmic}[1]
  \REQUIRE{$\mathbf{IN_{CRT}}(N,qLimbs)$, $\mathbf{TB_{CRT}}(np,qLimbs)$}
  \ENSURE{$\mathbf{OUT_{CRT}}(N,np)$}
  \FOR{$(i = 0; \; i < N; \; i \gets i + 1)$}
    \FOR{$(j = 0; \; j < np; \; j \gets j + 1)$}
    \State {$\mathbf{accum} \gets 0$}
    \FOR{$(k =0; \; k < qLimbs; k \gets k + 1)$}
    \State {$\mathbf{accum} \pluseq \mathbf{IN_{CRT}}[i][k] \times \mathbf{TB_{CRT}}[j][k]$}
    \ENDFOR
    \State{$\mathbf{OUT_{CRT}}[i][j] =\bmod(\mathbf{accum},p_j)$}
    \ENDFOR
  \ENDFOR 
\end{algorithmic}
\end{algorithm}

\textbf{\texttt{CRT}}
(Algo.~\ref{algorithm:crt}) takes $\mathbf{IN_{CRT}} (N, qLimbs)$ 
representing $N$ $\log{q}$-bit BigInt numbers % as input
and produces
$\mathbf{OUT_{CRT}} (N, np)$, the result of modulo operation on each BigInt 
with $np$ different primes $\{p_j|0\leq j < np\}$.
The operation consists of two stages: 
(1) computing matrix-matrix mul of $\mathbf{IN_{CRT}}$ with 
$\mathbf{TB_{CRT}}^\intercal$ and (2) applying modulo operations to each
output element.

We first explain how to perform a modulo operation on a BigInt with 
a modulus smaller than $\beta$.
A BigInt $A$ is expressed by $qLimbs$ $\log{\beta}$-bit words, 
i.e. $\sum_{k=0}^{qLimbs-1}a_k \cdot \beta^k$ where  $\{a_k | 0 \leq k < qLimbs\}$.
Then the modulo operation on the BigInt is as follows:
\begin{align*}
  \bmod(A,p_j) &= \bmod(\textstyle\sum_{k=0}^{{qLimbs}-1}a_k \cdot \beta^k,p_j)\\
  &= \bmod(\textstyle\sum_{k=0}^{qLimbs-1}a_k \cdot \bmod(\beta^k,p_j),p_j)
\end{align*}
Here because $\beta$ and $p_j$ are independent of the input, 
HEAAN precomputes $\mathbf{TB_{CRT}}$, $\bmod(\beta^k,p_j)$ for all $k$ and $j$. 
Therefore, $\sum_{k=0}^{qLimbs-1}a_k \cdot \bmod(\beta^k,p_j)$ is performed 
by multiplying $\mathbf{IN_{CRT}}$ and $\mathbf{TB_{CRT}}$.

\begin{algorithm}[tb!]
\caption{Shoup’s modular multiplication (ModMul)}
\small
\label{algorithm:shoup}
\begin{algorithmic}[1]
  \REQUIRE{$X, Y, p_j, Y_{Shoup}$}
  \ENSURE{$r = \bmod({X \times Y},p_j)$}
%  \State {$Q_{hi} = \text{mulhi}(X \times Yshoup)$}
  \State {$Qu_{hi} = (X \times Y_{Shoup}) \gg log\beta$}
  \State {$r = X \times Y - Qu_{hi} \times p_j$}
  \IF{$r > p_j$}
  \State{$r = r-p_j$}
  \ENDIF
\end{algorithmic}
\end{algorithm}

We can exploit Shoup's modular mul (Shoup's ModMul~\cite{ieee-2019-fpga}) 
for the modulo operation in line 6 of Algo.~\ref{algorithm:crt}. 
Shoup's ModMul (Algo.~\ref{algorithm:shoup}) computes 
$\bmod(X \cdot Y, p_j)$ with 3 muls and a single correction step
if a value $Y_{Shoup}$ ($=\lfloor\textstyle\frac{Y \cdot \beta}{p_j} \rfloor$) is known in advance.
It replaces a costly division operation with relatively cheaper mul, 
comparison, and subtraction operations.
We apply the algorithm for the modular mul on $accum$ 
spanning up to 3 limbs ($accum_0+accum_1 \cdot \beta+accum_2 \cdot \beta^2$),
using precomputed $Y_{Shoup}$ values on $Y = \{1, \beta, \beta^2\}$.
The operations of \texttt{CRT}
can be performed in parallel for each coefficient (total $N$) and each prime 
number (total $np$).

\begin{algorithm}[tb!]
\caption{NTT}
\small
\label{algorithm:ntt}
\begin{algorithmic}[1]
  \REQUIRE{$\mathbf{IN_{NTT}}(np,N) \gets \mathbf{OUT_{CRT}}^\intercal, \mathbf{TB_W}(np,N)$}
  \ENSURE{$\mathbf{OUT_{NTT}}(np,N)$}
	\FOR{$(i = 0; \; i < np; \; i \gets i+1)$}
  \State {$t =  N$}
  \State {$\mathbf{IN} = \mathbf{IN_{NTT}}[i]$}
  \State {$\mathbf{TBW} = \mathbf{TB_{W}}[i]$}
  \FOR{$(m = 1; \; m < N; \; m \gets m \times 2)$}
	  \State {$t = t \mathrel{/} 2$}
    \FOR{$(j = 0; \; j < m; \; j \gets j+1)$}
    \FOR{$(k = j \times 2t; \; k < j \times 2t + t; \; k \gets k+1)$}
      \State{$butt(\mathbf{IN}[k],\mathbf{IN}[k+t],p_i,\mathbf{TBW}[m+j])$}
    \ENDFOR
    \ENDFOR
	\ENDFOR
  \State {$\mathbf{OUT_{NTT}}[i] = \mathbf{IN}$}
  \ENDFOR
\end{algorithmic}
\end{algorithm}

\begin{algorithm}[tb!]
\caption{butt}
\small
\label{algorithm:butt}
\begin{algorithmic}[1]
  \REQUIRE{$A, B, p_i, W$}
  \ENSURE{$A,B$}
  \State {$U = \bmod(B \times  W,p_i)$}
	\State {$B = A \mathrel{-} U$}
	\State {$A = A \mathrel{+} U$}
\end{algorithmic}
\end{algorithm}

\textbf{\texttt{NTT}} implements Cooley-Tukey algorithm~\cite{mc-1965-ntt}, 
which recursively divides an $N$-point FFT to $k$ $N/k$-point FFTs 
and combines their results (called \emph{radix-k} FFT).
An exemplar radix-2 NTT in Algo.~\ref{algorithm:ntt} 
takes a matrix $\mathbf{IN_{NTT}} (np, N)$ as an input
and performs a butterfly algorithm \texttt{butt}.
It uses a precomputed table ($\mathbf{TB_{W}}$) of powers of the 2$N$-th root of
unity for all $np$ prime numbers.
For each prime, \texttt{butt} (Algo.~\ref{algorithm:butt}) is called 
log$N \cdot \frac{N}{2}$ times.
As \texttt{butt} requires modular mul, it also 
uses Shoup's ModMul as was done in \texttt{CRT}.

\textbf{\texttt{iNTT}} is slightly different from \texttt{NTT}.
It has a different loop order, calls inverse butterfly (\texttt{ibutt})
instead of \texttt{butt}, deals with a different precomputed table 
(consisting of the inverse of powers of the primitive root of 
unity ($\mathbf{TB_{invW}}$)), and finally divides each element by $N$.
However, except for the last element-wise division by $N$, \texttt{iNTT}
is symmetric to \texttt{NTT} in terms of the number and the kind of operations.
Both \texttt{NTT} and \texttt{iNTT} are completely parallelizable for each prime number.

\begin{algorithm}[tb!]
\caption{iCRT}
\small
\label{algorithm:icrt}
\begin{algorithmic}[1]
  \REQUIRE {$\mathbf{IN_{iCRT}}(np,N) \gets \mathbf{OUT_{iNTT}}(np,N)$}
	\Statex {$\mathbf{TB_{invP}}(np), \mathbf{TB_{Pdivp}}(np, PLimbs)$}
  \ENSURE{$\mathbf{OUT_{iCRT}}(N,np)$}
  \FOR{$(i = 0; \; i < N; \; i \gets i+1)$}
    \FOR{$(j = 0; \; j < np; \; j \gets j+1)$}
      \State {$\mathbf{temp}[j][i] = \bmod(\mathbf{IN_{iCRT}}[j][i]\times \mathbf{TB_{invP}}[j], p_j)$} 
		\ENDFOR
	\ENDFOR
  \FOR{$(i = 0; \; i < N; \; i \gets i+1)$}
	  \State{$\mathbf{accum} = 0$}
      \FOR{$(j = 0; \; j < np; \; j \gets j+1)$}
        \FOR{$(k = 0; \; k < PLimbs; \; k \gets k+1)$}
          \State{$\mathbf{accum} += \mathbf{temp}[j][i] \times \mathbf{TB_{Pdivp}}[j][k] \times \beta^{k}$}
        \ENDFOR  
	    \ENDFOR
    \State {$\mathbf{OUT_{iCRT}}[i] = \bmod(\bmod(\mathbf{accum},\mathbf{P}), \mathbf{q})$}
	\ENDFOR
\end{algorithmic}
\vspace{-0.05in}
\end{algorithm}

\begin{table*}[tb!]
  \centering
  \caption{The number of arithmetic operations and computational complexity of major functions of HE Mul.}
  \small
  \renewcommand{\arraystretch}{1.05}
  \begin{tabular}{lcccc}
    \toprule
    & CRT & NTT & iNTT & iCRT \\
    \midrule
    Multiplication            & $N\times qLimbs\times np$ &-                          &-                                     & $N\times np\times PLimbs$ \\ 
    Modular mul    & $N\times np$              & $np\times N/2\times \log N$ & $np\times (N/2\times \log N$ + $N)$ & $2\times N\times np$ \\
    ADC (add with carry) & $N\times qLimbs\times np$ &-                             &-                                    & $N\times np\times PLimbs$ \\
    Add, Sub                  &-                           & $np\times N\times \log N$   & $np\times N\times \log N$           &- \\
    \midrule
    Computation complexity    & $\mathcal{O}(N\times qLimbs\times np)$ & $\mathcal{O}(N\times logN\times np)$ & $\mathcal{O}(N\times logN\times np)$ & $\mathcal{O}(N\times np\times PLimbs)$ \\
    \bottomrule
  \end{tabular}
  \label{tab:num-operations}
\end{table*}

\begin{table}[tb!]
  \centering
  \caption{Input and precomputed data size of major functions of HE Mul.
  The unit of data size is $\beta$. Precomputed data for Shoup's ModMul
  is excluded.}
  \small
  \begin{tabular}{lccc}
    \toprule
    & CRT & NTT \& iNTT & iCRT \\
    \midrule
    Input data       & $N\times qLimbs$  & $N\times np$ & $N\times np$ \\
    Precomputed  & \multirow{2}*{$np\times qLimbs$} & \multirow{2}*{$N\times np$} & $np,$ \\
    data & & & $np\times PLimbs$ \\
    \bottomrule
  \end{tabular}
  \label{tab:data-size}
\end{table}

\textbf{\texttt{iCRT}} converts the matrix $\mathbf{OUT_{iNTT}} (np, N)$,
where each element is a remainder smaller than $\beta$,
back to $N$ $\log{q}$-bit BigInts (see Algo.~\ref{algorithm:icrt}).
It starts with (1) the Hadamard product between an input matrix and
a precomputed table $\mathbf{TB_{invP}}$ whose elements are modular inverses 
of $\sfrac{P}{p_j}$ for all $p_j$, followed by an element-wise modular mul 
with each $p_j$.
Shoup's ModMul can also be applied here for 
efficient modular mul.
(2) Then, each output element of (1), a scalar value,
is multiplied by a BigInt $\sfrac{P}{p_j}$ according to its $j$ and accumulated to a temporary 
BigInt $accum$.
Here each $\sfrac{P}{p_j}$ is precomputed and stored in  
table $\mathbf{TB_{Pdivp}} (np, PLimbs)$, where $PLimbs=\max\limits_{j}(\log({P/p_j})/\log{\beta})$.
(3) Finally, a reduction of $accum$ modulo $P$ and $q$ ($q \cdot Q$ in Region 2) is performed.

We summarize 
the number of arithmetic operations needed for each major function 
in Table~\ref{tab:num-operations} and
the size of input and precomputed data for each function in 
Table~\ref{tab:data-size}.

\section{Architecture-aware Optimizations to Maximize HE Mul performance on CPUs and GPUs}
\label{sec:contribution-3}

Previous HE studies~\cite{arxiv-2019-heax,ieee-2019-fpga} pursued proposing
new hardware architectures (e.g., through FPGA implementation) for performance
improvement. 
By contrast, we first improve the performance of HE
by utilizing the most popular computation platforms, CPUs and GPUs, 
which are already equipped with hundreds to thousands of ALUs.

All the major functions (\texttt{CRT}, \texttt{iCRT}, \texttt{NTT},
and \texttt{iNTT}) of HE Mul of HEAAN have massive parallelism
that can be exploited by CPUs and GPUs. 
All the residual numbers ($N \times np$) can be computed in parallel 
on \texttt{CRT}.
\texttt{NTT} and \texttt{iNTT} perform $np$ independent transformations
and leverage the algorithmic 
optimization of FFT, where $N/k$ groups can be computed in parallel at
each individual stage during FFT. 
Henceforth, we identify key challenges and solutions we devise 
in accelerating HE Mul on CPUs and GPUs.

\begin{algorithm}[tb!]
\caption{iCRT algorithm in the matrix-matrix mul form.
  The first three lines are the same as Algo.~\ref{algorithm:icrt}.} % five lines with endfor
\small
\label{algorithm:icrt_matrix}
\begin{algorithmic}[1]
		\setcounter{ALG@line}{3} % 5 with endfor
		\Statex {...}\
    \FOR{$(i = 0; \; i < N; \; i \gets i+1)$}
	  \State{$\mathbf{accum} = 0$}
      \FOR{$(k = 0; \; k < PLimbs; \; k \gets k+1)$}
			\State {$\mathbf{accum_{small}}=0$}
        \FOR{$(j = 0; \; j < np; \; j \gets j+1)$}
          \State{$\mathbf{accum_{small}} += \mathbf{temp}[j][i] \times \mathbf{TB_{Pdivp}}[j][k]$}
        \ENDFOR
				\State{$\mathbf{accum} = \mathbf{accum} + \mathbf{accum_{small}} \times \beta^{k}$}
	    \ENDFOR
    \State {$\mathbf{OUT_{iCRT}}[i] = \bmod(\bmod(\mathbf{accum},\mathbf{P}), \mathbf{q})$}
	\ENDFOR
\end{algorithmic}
\end{algorithm}
\vspace{-0.1in}

\subsection{Loop reordering to expose massive parallelism in iCRT}
\label{sec:icrt-loop-reordering}

\texttt{iCRT} recombines the residual numbers into the integers of size log$q$
for each coefficient of the resulting ciphertext and hence it might be regarded
that the degree of parallelism is smaller than \texttt{CRT} ($N$ vs. $N\!\cdot\!np$).
However, we discover that the limited $N$-degree parallelism can be expanded 
to $N\!\cdot\!PLimbs$-degree parallelism by reordering 
two loops in \texttt{iCRT} (line 6 and 7 in Algo.~\ref{algorithm:icrt}); % line 8 and 9 with endfor
the modified algorithm is shown in Algo.~\ref{algorithm:icrt_matrix}.
After reordering, the sequence of original mul between a scalar and a BigInt
now becomes a matrix-matrix mul between a $temp$ matrix 
and a $\mathbf{TB_{Pdivp}}$ matrix (line 9 in Algo.~\ref{algorithm:icrt_matrix}). % line 11 with endfor
Then, \texttt{iCRT} should be modified such that 
the partial sum in the inner-most loop is accumulated into $accum_{small}$ 
(double or triple words), rather than $accum$ (BigInt), 
which is aggregated to $accum$ at the end of the loop.

The range of modulus and $np$ determine whether 
$accum_{small}$ should be a double word or a triple word;
with our representative parameters 
where $\beta = 2^{64}$, using a double word is sufficient.
With our loop reordering, the resulting matrix-matrix mul exposes
a massive parallelism of degree $N\!\cdot\!PLimbs$ in \texttt{iCRT}, providing
abundant parallelization opportunities to contemporary hardware platforms.

\subsection{Accelerating HE Mul on CPUs}
\label{sec:cpus}

The strategies a modern CPU takes to exploit parallelism from an application are 
twofold: populating (1) multiple cores and (2) ALUs supporting short-SIMD instructions 
within each core.
For example, the Skylake-based Intel Xeon CPU we use has 24 cores per 
socket and each core support AVX-512 instructions, each executing
eight 64-bit integer operations~\cite{micro-2017-skylake}, which could
lead to over 100$\times$ performance improvement compared to the baseline
implementation not exploiting these parallelism. %intra-core and inter-core parallelism. 
To achieve a near-maximum performance CPU provides, we exploit intra-core 
parallelism via utilizing AVX-512 instructions and inter-core parallelism
via multi-threading.
We carefully distribute operations to multiple threads and AVX-512 SIMD lanes 
per thread to minimize performance degradation by an inferior cache
performance caused by poor data access patterns; in this, 
whether an input matrix follows a column- or a row-major order can vastly 
affect performance.

During \texttt{CRT}, a CPU thread takes responsibility of a portion of the
$N$ coefficients (line 1 in Algo.~\ref{algorithm:crt}),
whereas each lane of an AVX-512 port performs operations on different prime
numbers (line 2 in Algo.~\ref{algorithm:crt}).
During \texttt{NTT} and \texttt{iNTT}, % on the other hand,
a thread does
its job on a portion of the prime numbers, 
(line 1 in Algo.~\ref{algorithm:ntt}),
whereas each lane of AVX-512 computes a part of the coefficients 
(line 8 in Algo.~\ref{algorithm:ntt}).
In case of \texttt{iCRT}, we take different approaches for the two
iteration phases.
During the first phase (line 1 to 3 in Algo.~\ref{algorithm:icrt}), each % line 1 to 5 with endfor
thread and a AVX-512 lane performs computation on a part of the $N$
coefficients. 
During the second iteration phase (line 4 to 11 in Algo.~\ref{algorithm:icrt_matrix}) % line 6 to 13 with endfor
each thread also computes on a part of the coefficients (line 4), % in Algo.~\ref{algorithm:icrt}), % line 6 with endfor
but each lane of an AVX-512 port is on different $k$, the positional index on the limbs
of $\sfrac{P}{p_j}$ (line 6). % in Algo.~\ref{algorithm:icrt}). % line 8 with endfor

Although the reference HEAAN library~\cite{repo-2019-heaan} supports
multi-threading and exploits the same parallelism type on \texttt{NTT},
\texttt{iNTT}, and \texttt{iCRT} with our work,
(in terms of distributing works to threads; the reference HEAAN does
not implements SIMD instructions) \texttt{CRT} takes a different strategy.

\noindent
\textbf{Emulating arithmetic operations:}
However, AVX-512 does not support parallel 64-bit mul and 64-bit
ADC (addition with carry) yet. 
Therefore, 64-bit mul is emulated with four parallel 32-bit 
mul, five 64-bit add, and five 64-bit shift instructions. 
Also, one 64-bit compare and one additional 64-bit add are required to
handle carry per addition. 
This emulation narrows the performance gap between the reference HEAAN %baseline implementation
and the AVX-512 implementation. 
To further improve performance under this constraint, we modify Shoup's ModMul 
as follows.

The original Shoup's ModMul algorithm requires three operations: 
one 64-bit mulhi operation to compute $Qu_{hi}$, and two 64-bit mullo operations to 
compute $Qu_{hi} \cdot p_j$ and $X \cdot Y$, 
where $Qu$ is an estimation of a quotient, and 64-bit mulhi (mullo) returns the upper 
(lower) 64-bit of a mul result which is 128-bit long.
A single 64-bit mulhi operation can be emulated with four 32-bit mul 
($hi \cdot hi$, $hi \cdot lo$, $lo \cdot hi$, $lo \cdot lo$), four 64-bit add, and five 64-bit shift operations.
In this case, the estimated remainder can be either $r$ or $r+p_j$, which lies in a range 
of $[0, 2p_j)$.

However, one of the 32-bit mul ($lo \cdot lo$) for emulating 64-bit mulhi is used only 
for computing a carry from low 64-bit of mul.
We can remove this $lo \cdot lo$ mul if the carry is ignored, and produce an 
approximated 64-bit mulhi ($Qu_{hi'}$) with only three mul instead of four.
By applying this optimization, the estimated remainder lies in a range of $[0, 4p_j)$.
As the upper bound of a remainder grows to $4p_j$, one more correction step (conditional 
subtraction) is needed, but the number of more expensive operations are reduced: 
one 32-bit mul, two 64-bit add, and one 64-bit shift instruction.

\noindent
\textbf{Matrix transposition in \texttt{iCRT}:}
SIMD instructions might lead to a poor cache utilization if they access a matrix
storing elements in a row-major order by column direction (or vice versa), as 
this demands multiple cache lines at once.
The resulting performance degradation is more prominent when the access stride is
too large for a hardware prefetcher to be effective.
\texttt{iCRT} experiences this issue because the matrix-matrix mul
(line 9 in Algo.~\ref{algorithm:icrt_matrix}) accesses the $temp$ matrix storing % line 11 with endfor
elements in a row-major order by column direction.
We implicitly transfer the $temp$ matrix using the \emph{scatter} instructions
in AVX-512 to address this issue.

\noindent
\textbf{Reducing the size of $\beta$ to $2^{32}$:}
We can remove the overhead of emulating 64-bit operations by using $\beta$ of
$2^{32}$ instead of $2^{64}$ as AVX-512 naturally supports 32-bit 
mul and ADC. % operations.
Using the smaller beta changes the number of instructions for mul and 
ADC to one, and reduces the number of instructions for modular mul to 
below the half.
However, it also has shortcomings that $qLimbs$ should roughly be doubled to 
express numbers whose size is up to $q$. $np$ also needs to be doubled
because the upper bound of each prime number is $\beta$.
Larger $qLimbs$ and $np$ imply bigger precomputed tables and more iterations
for \texttt{CRT}, \texttt{NTT}, \texttt{iNTT}, and \texttt{iCRT}.
It also increases the number of operations other than mul, modular
mul, and ADC.
A preliminary implementation shows no improvement in performance by using the
small $\beta$ for CPUs.

\subsection{Accelerating HE Mul on GPUs}
\label{sec:gpus}

Modern GPUs such as Volta~\cite{nvidia-2017-volta}
and Turing~\cite{nvidia-2018-turing} have as many integer (INT32) processing
units as single-precision floating-point (FP32) processing units, resulting in
the multiply-accumulate throughput for INT32 numbers the same as that for FP32
numbers.
Such massive throughput of integer operations makes GPU an attractive candidate
for accelerating HE operations.
We identify the following key points in fully exploiting the performance 
potential of GPUs when accelerating HE operations.

\noindent
\textbf{Parallelization strategies:}
CUDA programming model~\cite{nvidia-2019-cudac} for GPU has the 
following hierarchical structure of threads: multiple GPU threads are
grouped to form a \emph{thread block} and multiple thread blocks 
comprise a \emph{grid}.
A thread block is allocated to one of Streaming Multiprocessors (SMs).
The threads in a thread block share the resources (e.g., shared memory) of
the SM.
Each thread block in a grid is allocated to each SM in a round-robin fashion, 
and the number of thread blocks (\emph{grid dimension}) and the number of
threads in a block (\emph{block dimension}) are configured
at each GPU kernel launch.

The basic parallelization strategy is to assign each independently computable 
output element in a function to a GPU thread.
We launch $N \cdot np$ threads for \texttt{CRT} so that each thread 
computes one output element, which is a residue. 
\texttt{NTT} and \texttt{iNTT} launch $\sfrac{N}{2} \cdot np$ threads each, where
one thread performs one butterfly operation (Algo.~\ref{algorithm:butt})
for each butterfly step using a simple radix-2 iterative NTT 
algorithm~\cite{jsc-2014-harvey} that is also used for CPU implementation.

In case of \texttt{iCRT}, a na\"ive parallelism strategy uses $N$-degree
parallelism where one thread handles one output BigInt type coefficient.
Prior studies~\cite{iccis-2015-cuhe, iacr-2018-fvgpu} took the same strategy. 
However, by changing the loop order as described above, we transform its
core operation into a matrix-matrix mul operation,
thereby taking advantage of $N \cdot PLimbs$-degree parallelism to maximize TLP. 
We can also exploit various optimization strategies developed for matrix-matrix
mul in GPU (e.g., tiling principles in NVIDIA's cutlass 
library~\cite{nvidia-2013-cutlass}).

\noindent
\textbf{64-bit emulation vs. 32-bit word:}
As opposed to most CPUs that natively support 64-bit words, modern GPUs natively
support 32-bit words and emulate 64-bit integer operations.
To avoid the overhead of 64-bit emulation (whose throughput is more than an 
order of magnitude lower than that of the 32-bit counterpart), prior studies
accelerating HE on GPU~\cite{iacr-2018-fvgpu, iccis-2015-cuhe} use 32-bit words
($\beta = 2^{32}$) and the prime numbers smaller than $\beta$. 
We also use 32-bit words and operations.
Another advantage of using 32-bit words on GPU is that the operations with carry-in and 
carry-out can be used without emulation;
modern NVIDIA GPUs support carry operations (e.g., addc, subc, and madc
in assembly-like virtual ISA, PTX~\cite{nvidia-2019-ptx} 
where the operations are called \emph{extended-precision integer arithmetic instructions}).
The throughput of these instructions is the same as
the instructions without carry in recent GPU 
architectures~\cite{nvidia-2017-volta, nvidia-2018-turing},
enabling efficient computation on large integers without emulation.

\noindent
\textbf{Different strategies for BigInt modulo in \texttt{CRT}:}
A na\"ive BigInt modulo is done by repetitive log$\beta$-bit shift, add, and modulo operation; 
for example, cuHE~\cite{iccis-2015-cuhe}, takes this approach.
By contrast, HEAAN accumulates the result of modulo on each $a_k \cdot \beta^k$ 
using a precomputed table in \texttt{CRT}.
In this case, the BigInt variable $accum$ (line 3 in Algo.~\ref{algorithm:crt}) 
can span two or three words depending on $np$ and the size of each prime number.
In the CPU implementation, $accum$ is guaranteed to span two words when using the
representative parameters specified in Table~\ref{tab:experiment-parameters},
as an overflow does not happen for $np \leq 2^{64-58}=64$ with prime numbers smaller
than $2^{58}$.
To guarantee $accum$ to be two-word long, we use $2^{57}$ as
a lower bound of a prime for AVX-512 implementation instead of $2^{59}$
which is the default value of the reference HEAAN.
However in GPU with a 32-bit $\beta$, with primes smaller than 
$2^{30}$,\footnote{We used $2^{27}$ as a lower bound of a prime when 
$\beta = 2^{32}$.  If the lower bound is excessively reduced (e.g., using 
$2^{21}$) to avoid any overflow, the overhead from a growing $np$ becomes
too high, negating the advantage of removing overflows.}
only up to 4 ($= 2^{32-30}$) accumulation is allowed to guarantee that the 
overflow does not happen, which is nearly impossible as $np$ is 90 or higher
when using the representative parameters.

To prevent the overflow, one might 
(1) use three-word $accum$ with an additional ADC operation added in the 
inner-most loop to avoid expensive modulo operations, or
(2) do modulo operations intermittently in the inner-most loop (e.g., for
every 4 accumulation in our case),
to ensure that $accum$ spans only two words.
We compare these strategies in Section~\ref{sec:evaluation}.

\noindent
\textbf{Per-thread storage for accumulation in \texttt{iCRT}:}
The baseline implementation of \texttt{iCRT} with N-degree parallelism
allocates a BigInt $accum$ (line 8 in Algo.~\ref{algorithm:icrt}) % line 10 with endfor
as a long array, in a per-GPU-thread manner.
If $accum$ is not carefully allocated to fast storage, 
frequent cache thrashing might occur, leading to a significant performance degradation.
The latest NVIDIA GPUs~\cite{nvidia-2017-volta, nvidia-2018-turing} have 
a variety of storage types including register, L1 cache, L2 cache,
device memory, and read-only constant memory.
In the algorithm of original \texttt{iCRT} (Algo.~\ref{algorithm:icrt}), $temp[j][i]$ is 
stored in register memory, 
so that it can be loaded quickly in a single cycle. 
On the other hand, because $accum$ is declared as a thread-local array
and also it is dynamically indexed in the algorithm (e.g., used as $accum[idx]$ 
where $idx$ is a variable), 
it is not stored in register, which is the fastest storage on GPU.
Instead, CUDA compiler stores it in global memory and caches into L1 and L2 (in CUDA programming 
model this is called \emph{local memory}).

However, heavily using local memory can lead to cache thrashing when the grid dimension 
and block dimension increase, 
leading to a number of threads competing for cache, degrading overall performance.
In order to mitigate the cache miss penalty, we suggest two different optimizations when 
using $N$ parallelism in \texttt{iCRT}: 
(1) using fewer threads by simply reducing block dimension and grid dimension, using the 
grid-stride loop method~\cite{nvidia-2013-gridstride}, or
(2) pinning each $accum$ array in L1 cache through allocating the array in shared memory; 
this is possible as the shared memory shares capacity with the L1 unified cache.
(2) is similar to the implementation of CRT kernel in 
cuHE~\cite{iccis-2015-cuhe},
which uses the shared memory for storing thread-local arrays.
We compare the methods on cuHE's iCRT kernel~\cite{iccis-2015-cuhe}
which implements Algo.~\ref{algorithm:icrt} with $N$ parallelism,
along with loop reordering with $N \cdot PLimbs$ parallelism which is explained
in Section~\ref{sec:icrt-loop-reordering}.

\noindent
\textbf{High-radix \texttt{NTT} (\texttt{iNTT}):}
Radix-2 \texttt{NTT} is memory-bound; GPU reads and writes a
large input $\mathbf{IN_{NTT}}(np,N)$ (dozens of megabytes with typical $np$ and
$N$ values specified in Table~\ref{table:formatting}, exceeding the size of L1
and L2 caches of GPU) by $\log_2{N}$ times.
At each butt function in Algo.~\ref{algorithm:ntt}, a GPU thread reads two values 
of $\mathbf{IN}$ from the device memory and writes two output values back to
the device memory.

Using high radix \texttt{NTT} (radix-$k$ with $k > 2$) can mitigate the 
memory bandwidth bottleneck because each GPU thread reads and writes $k$ values within
$\mathbf{IN}$ in the butt operation performing $k$-point NTT.
It changes the number of transferring $\mathbf{IN}$ 
from $\log_2{N}$ above to $\log_k{N}$, reducing the number of main memory 
accesses needed for \texttt{NTT}. 
However, increasing the radix is not always beneficial because, as each thread
takes more than two inputs, register pressure on each GPU thread increases.
Using registers more than the register file size of a streaming multiprocessor
causes register spilling to local memory, leading to performance degradation 
with additional data loads from the main memory.
Given the constraint, we use an appropriate size of radix (radix-32) to 
alleviate the high pressure of main memory bandwidth and conduct a sensitivity 
study in Section~\ref{sec:evaluation}.

\begin{table}[tb!]
  \centering
  \caption{HEAAN parameter settings for CPU and GPU.}
  \small
  \vspace{0in}
  \begin{tabular}{lcc}
    \toprule
    Parameters & CPU (AVX-512) & GPU \\
    \midrule
    $N$ & \multicolumn{2}{c}{$2^{16}$} \\
    $Q$ \& $q$ & \multicolumn{2}{c}{$2^{1200}$} \\
    $\beta$ \& $qLimbs$ & $2^{64}$ \& 19 & $2^{32}$ \& 38 \\
    Prime number size & $2^{57} < p_i < 2^{60}$ & $2^{27} < p_i < 2^{30}$ \\
    $np$ (Region 1) & 42 (43) & 90 \\
    $np$ (Region 2) & 63 (64) & 134 \\
    \bottomrule
  \end{tabular}
  \vspace{-0.1in}
  \label{tab:experiment-parameters}
\end{table}

\section{Experimental Setup}
\label{sec:experimental-setup}

We compared the performance of the reference HEAAN~\cite{repo-2019-heaan},
our AVX-512 implementation with multi-threading, and GPU implementation. 
We used Intel Xeon CPU (Skylake-based Xeon Platinum 8160 operating at 2.1 GHz)
and NVIDIA GPU (Turing Titan RTX operating at 1.35 GHz).
The CPU system consists of 24 cores per socket and each core has two AVX-512 
FMA units, achieving a peak 64-bit integer performance of 1.61 TOPS per socket. 
Each socket has six memory channels, each equipped with DDR4-2666 DRAM modules. 
We did not use HyperThreading; the number of cores utilized was the same as the
number of CPU threads populated.
The GPU system consists of 72 streaming multiprocessors (SMs), 
each with 64 CUDA cores, performing up to 4,608 32-bit integer operations per cycle. 
The size of each L1 unified cache and L2 shared cache in the GPU is 128 KB (per SM) 
and 6 MB, respectively. 
Even if the CPU system has two CPU sockets, we only used one socket to compare it
with the GPU system with a single discrete GPU.

We tabulate the key parameters for HE Mul of HEAAN on CPU and GPU
in Table~\ref{tab:experiment-parameters}.
We measured the execution time of HE Mul, excluding time for
memory operations, such as malloc, free, and data transfers from host to 
the device for GPU.
We conducted each experiment 32 times and reported the average.

\begin{table*}[tb!]
  \centering
  \caption{Comparing the execution time of HE Mul among
  a single- and 24-thread reference HEAAN ($\textbf{Ref}$-1 and $\textbf{Ref}$-24),
  a 24-thread optimized AVX-512 implementation ($\textbf{AVX-MT}$-24), and an optimized GPU 
  implementation ($\textbf{GPU-CLH}$).}
  \small
  \begin{tabular}{l|S[table-number-alignment = center]|S[table-number-alignment = center]r
    |S[table-number-alignment = center]r|S[table-number-alignment = center]r}
    \toprule
    & \multicolumn{7}{c}{Execution time (ms) and [Relative speedup]} \\
    \midrule
    Function & \multicolumn{1}{c|}{$\textbf{Ref}$-1 (baseline)} & \multicolumn{2}{c|}{$\textbf{Ref}$-24} & \multicolumn{2}{c|}{$\textbf{AVX-MT}$-24} & \multicolumn{2}{c}{$\textbf{GPU-CLH}$} \\
    \midrule
    $\texttt{CRT}$   & 639.9  & 40.4  & [15.8$\times$] & 17.5  & [36.6$\times$]  & 4.1  & [156.1$\times$] \\
    $\texttt{NTT}$   & 1541.0 & 73.3  & [21.0$\times$] & 8.7   & [177.1$\times$] & 3.7  & [416.5$\times$] \\
    $\texttt{iNTT}$  & 584.6  & 28.3  & [20.7$\times$] & 10.2  & [57.3$\times$]  & 4.7 & [124.4$\times$]  \\
    $\texttt{iCRT}$  & 2126.7 & 130.1 & [16.3$\times$] & 45.4  & [46.8$\times$]  & 19.4 & [109.6$\times$] \\
    Extra & 215.7  & 73.1  & [3.0$\times$]  & 37.3  & [5.8$\times$]   & 6.2  & [34.8$\times$]  \\
    \midrule
    Total & 5108.0 & 345.3 & [14.8$\times$] & 119.1 & [\textbf{42.9$\times$}]  & 38.1 & [\textbf{134.1$\times$}] \\
    \bottomrule
  \end{tabular}
  \label{tab:eval-avx-gpu}
  \vspace{-0.08in}
\end{table*}

\section{Evaluation}
\label{sec:evaluation}

We evaluated the effectiveness of the proposed optimizations in accelerating
HEAAN mul by comparing against the performance of the reference HEAAN
($\textbf{Ref}$).
For CPU, our optimizations exploit both intra-core and inter-core parallelism.
We compared
the basic implementation utilizing AVX-512 instructions ($\textbf{AVX}$),
the one with the modified Shoup's ModMul on top of $\textbf{AVX}$ ($\textbf{AVX-M}$),
and the one transposing the $temp$ matrix on top of $\textbf{AVX-M}$ ($\textbf{AVX-MT}$).
In the basic GPU implementation ($\textbf{GPU}$), we adopted radix-2 iterative NTT
for \texttt{NTT} and \texttt{iNTT}.
We modified the CRT kernel of cuHE~\cite{iccis-2015-cuhe}, which only exploits
$N$-degree parallelism, to exploit $N$$\cdot$$np$-degree parallelism.
Also, we used the iCRT kernel of cuHE.
We compared $\textbf{GPU}$ with the followings:
the implementation optimizing \texttt{CRT} by using ADC instead of
intermittently conducting modulo operations ($\textbf{GPU-C}$),
the one adjusting the number of launching threads on top of $\textbf{GPU-C}$ ($\textbf{GPU-CT}$),
the one using shared memory to pin the arrays of each thread to L1 unified 
cache on top of $\textbf{GPU-C}$ ($\textbf{GPU-CP}$),
the one applying loop reordering (Algo.~\ref{algorithm:icrt_matrix})
to translate a majority of \texttt{iCRT}
computation into matrix-matrix mul to use $N$$\cdot$$PLimbs$-degree
parallelism on top of $\textbf{GPU-C}$ ($\textbf{GPU-CL}$),
and the one implemented with high-radix \texttt{NTT} and \texttt{iNTT} 
to reduce main memory accesses and utilize GPU’s computing power more efficiently
on top of $\textbf{GPU-CL}$ ($\textbf{GPU-CLH}$).

We made the following key observations.
\emph{First, exploiting the massive parallelism supported by modern CPUs and GPUs gives
even more than 100$\times$ performance improvement in HE Mul.}
Table~\ref{tab:eval-avx-gpu} shows the execution time and the relative speedup
of the CPU and GPU implementations after applying a series of architecture-aware
optimizations.
$\textbf{AVX-MT}$ and $\textbf{GPU-CLH}$, the implementations giving the best
performance for CPU and GPU, achieve 42.9$\times$ and 134.1$\times$
speedup, respectively, compared to the single-thread $\textbf{Ref}$. 
$\textbf{GPU-CLH}$ performs 4.3$\times$ and 2.3$\times$ better than $\textbf{AVX-MT}$
on \texttt{CRT} and \texttt{iCRT} thanks to more ALUs populated on GPU.
Also, by reducing the main memory accesses through increasing the radix, 
$\textbf{GPU-CLH}$ achieves 2.35$\times$ and 2.17$\times$ performance improvement 
in \texttt{NTT} and \texttt{iNTT}, respectively, compared to $\textbf{AVX-MT}$’s implementation.

\begin{figure*}[!tb]
  \center
  \subfloat[Execution time on single-thread]{\includegraphics[height=1.25in]{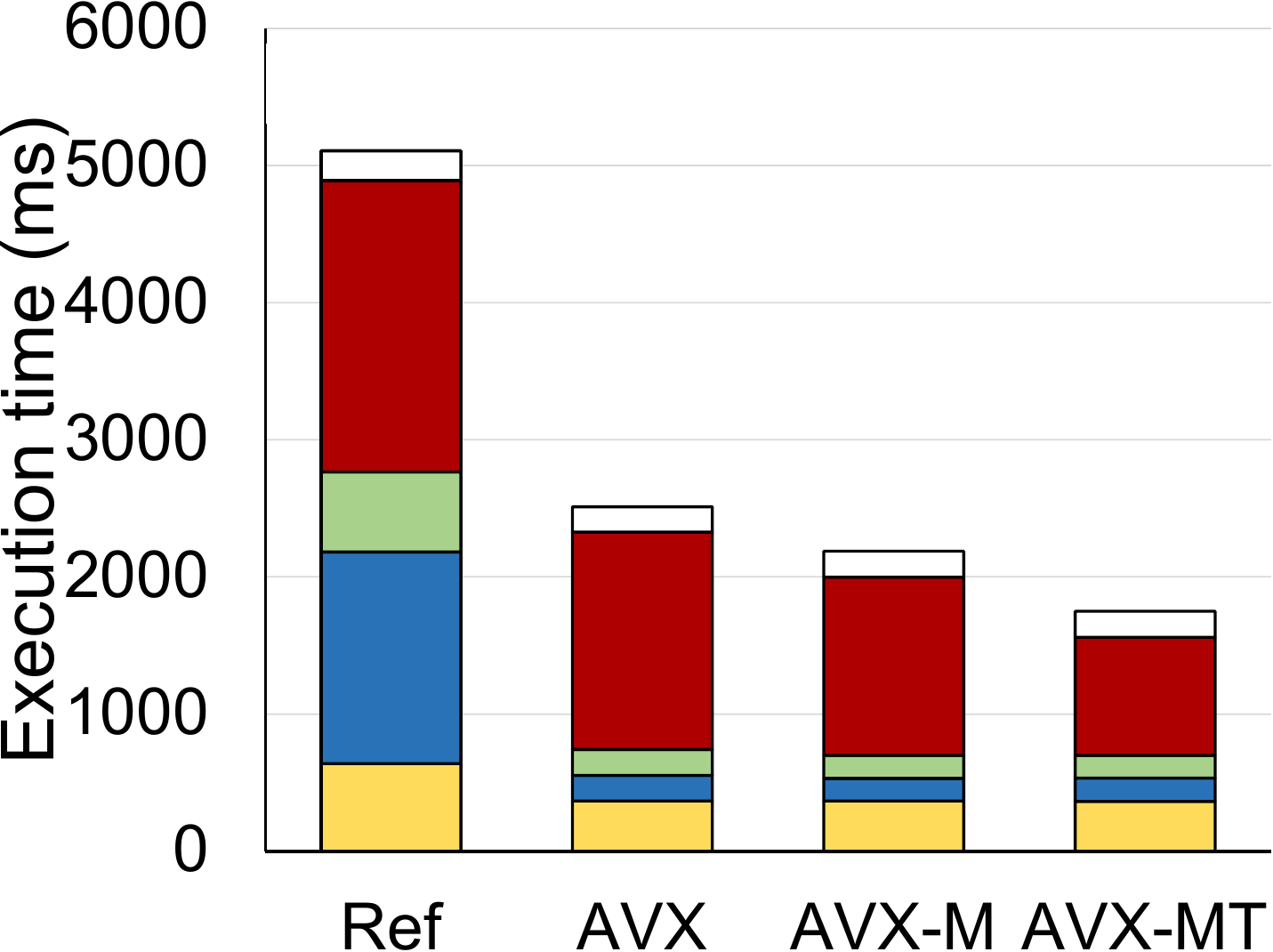}}
  \hspace{0.1in}
  \subfloat[Execution time on 24-threads]{\includegraphics[height=1.25in]{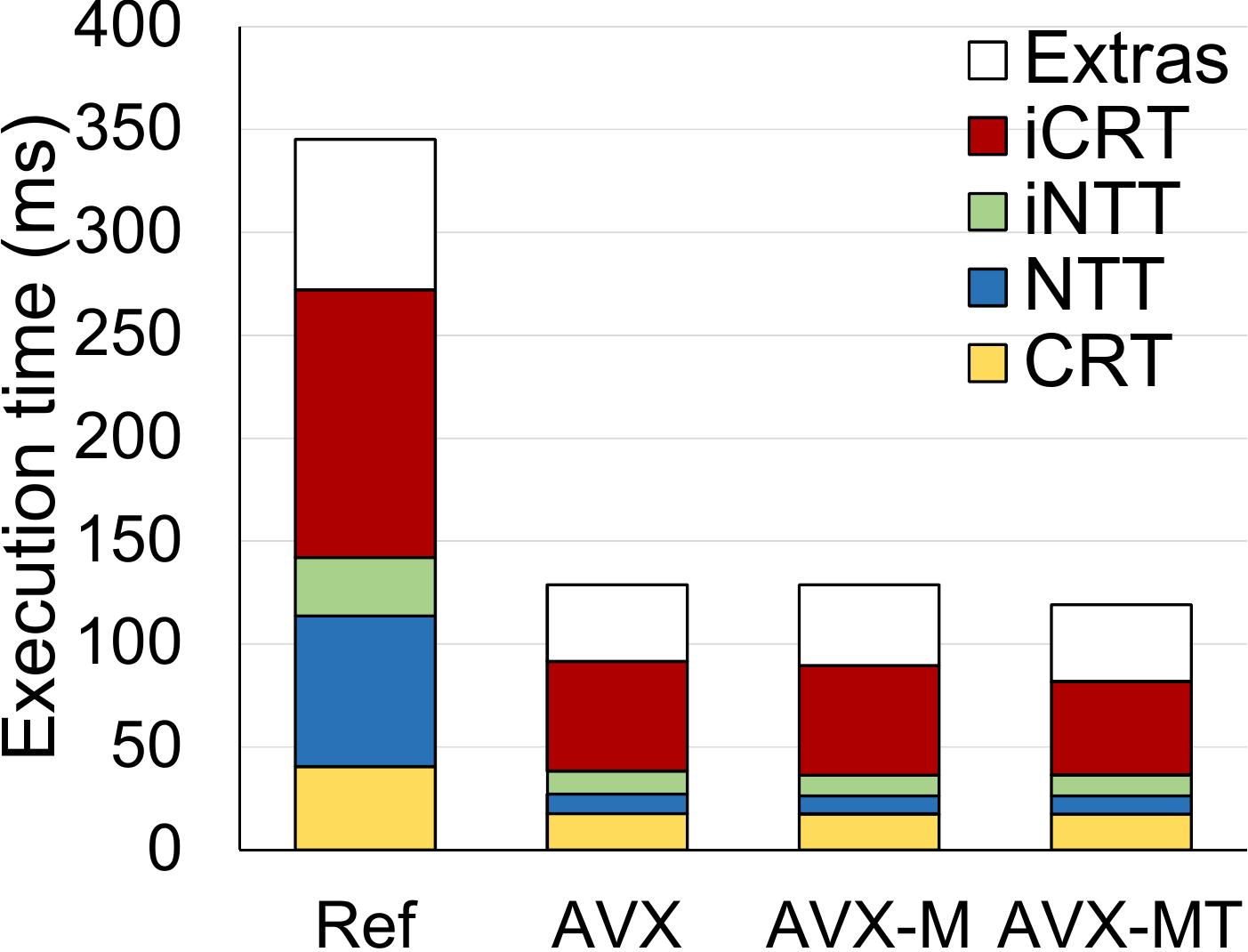}}
  \hspace{0.2in}
  \subfloat[Relative execution time and speedup per function]{\includegraphics[width=3.3in]{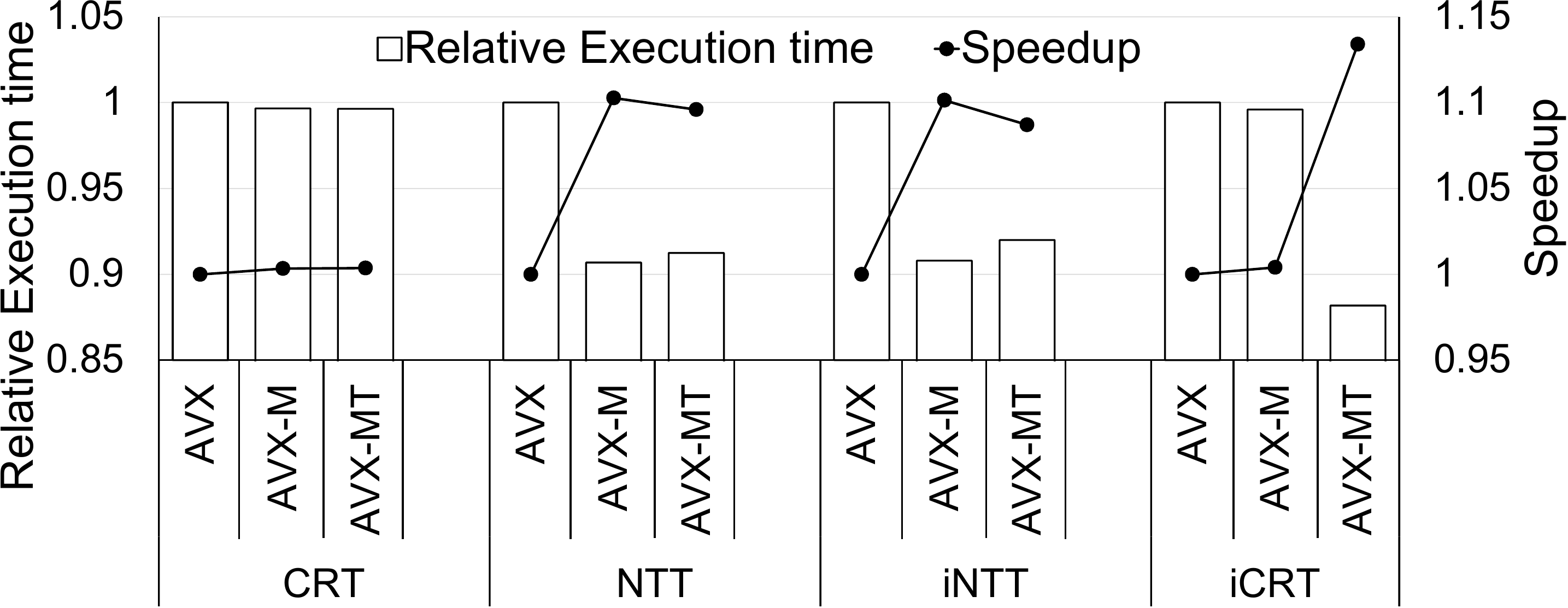}}
  \caption{
    Comparing HE Mul execution time among 
    $\textbf{Ref}$, AVX-512 implementation ($\textbf{AVX}$), and optimized $\textbf{AVX}$ 
    ($\textbf{AVX-M}$ and $\textbf{AVX-MT}$) when (a) one and (b)
    24 threads are utilized, and (c) per-function speedup when using 24 threads. 
  }
  \label{fig:eval-avx}
  \vspace{-0.08in}
\end{figure*}

\emph{Second, our CPU implementations are highly scalable across both intra-core
and inter-core dimensions.}
$\textbf{AVX}$ is effective regardless of the number of CPU threads populated,
providing 2.0$\times$ and 2.7$\times$ performance gain over $\textbf{Ref}$
when a single and 24 threads are utilized, respectively (see Figure~\ref{fig:eval-avx}(a) and (b)).
Among the primary functions, \texttt{NTT} is the best in scalability, leading
to 7.7$\times$ speedup for $\textbf{AVX}$ over $\textbf{Ref}$ when 24 threads
are populated.
Overall, $\textbf{AVX}$ experiences 19.4$\times$ speedup when the number of
populated threads increases from 1 to 24, exhibiting a better scalability 
than $\textbf{Ref}$ (14.8$\times$) because of the following reasons.
$\textbf{AVX}$ and $\textbf{Ref}$ exploit parallelism in different ways for
\texttt{CRT} as described in Section~\ref{sec:contribution-3}.
In $\textbf{Ref}$, each thread operates on different prime numbers 
($np$-degree parallelism) where $np$ is not large (e.g., 42 or 63), and
hence $\textbf{Ref}$ is more susceptible to a load imbalance across threads.
By contrast, each thread operates on different coefficients ($N$-degree 
parallelism) in $\textbf{AVX}$, exhibiting better scalability.
For \texttt{iCRT}, data accesses for the matrix occur in column direction
during matrix-matrix mul, causing its performance memory-bound because
hardware prefetching becomes ineffective.
However, with 24 threads being utilized, hardware prefetching hits more 
frequently because a thread might access the data in the adjacent columns
that are prefetched by other threads, leading to better performance.

The additional optimizations applied to the AVX-512 implementation are 
effective as well.
Figure~\ref{fig:eval-avx}(c) shows the impact of these optimizations
on each major function when 24 threads are used.
In $\textbf{AVX-M}$, both \texttt{NTT} and \texttt{iNTT} are 10\% faster than 
$\textbf{AVX}$ because these functions compute modular mul frequently.
\texttt{iCRT} experiences a 18\% speedup in $\textbf{AVX-MT}$ compared to 
$\textbf{AVX-M}$ because the matrix transposition alleviates the memory-bound issue.

\begin{figure*}[!tb]
  \center
  \subfloat[HE Mul execution time breakdown]{\includegraphics[height=1.2in]{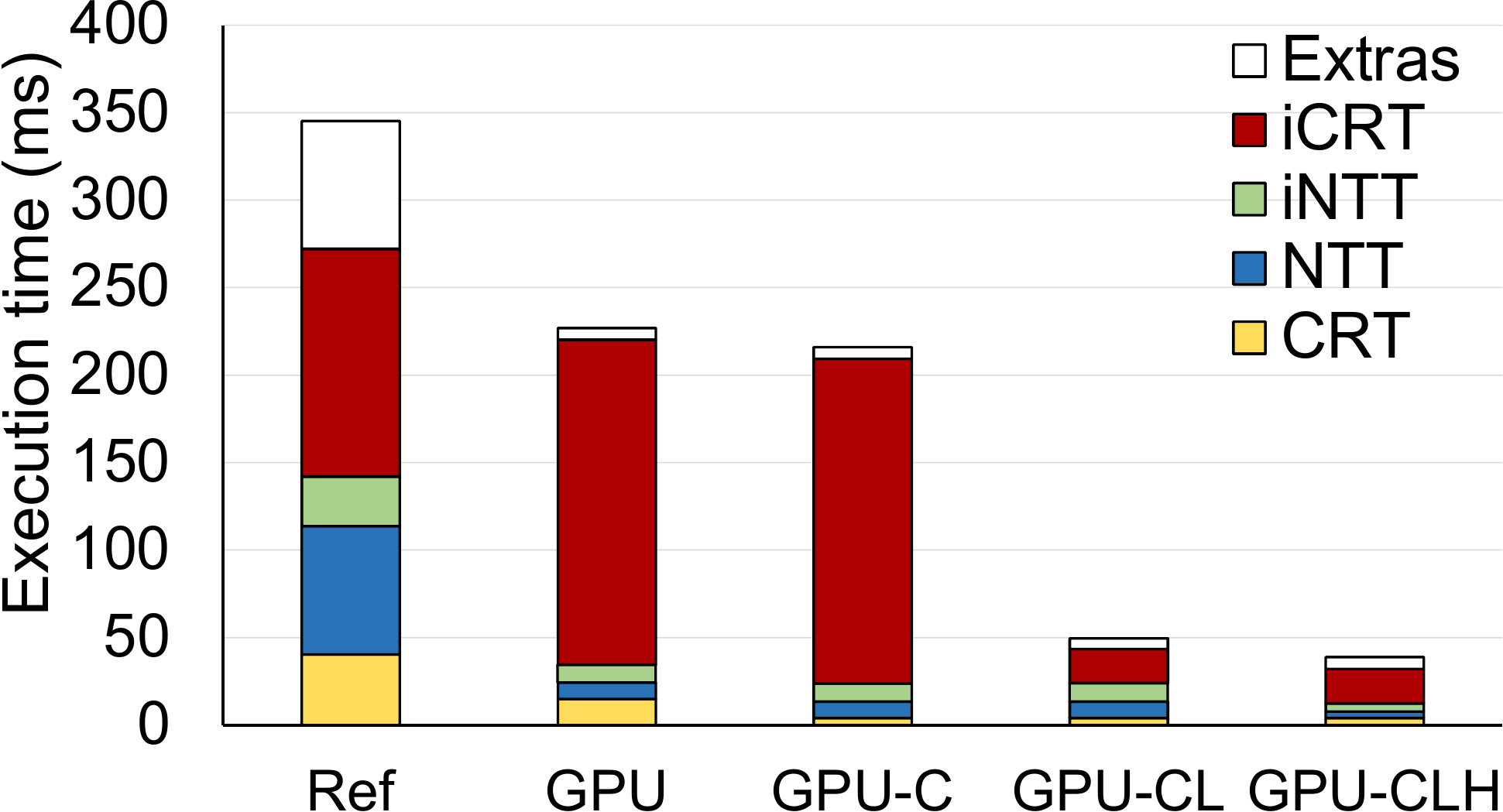}}
  \hspace{0.05in}
  \subfloat[CRT optimization]{\includegraphics[height=1.2in]{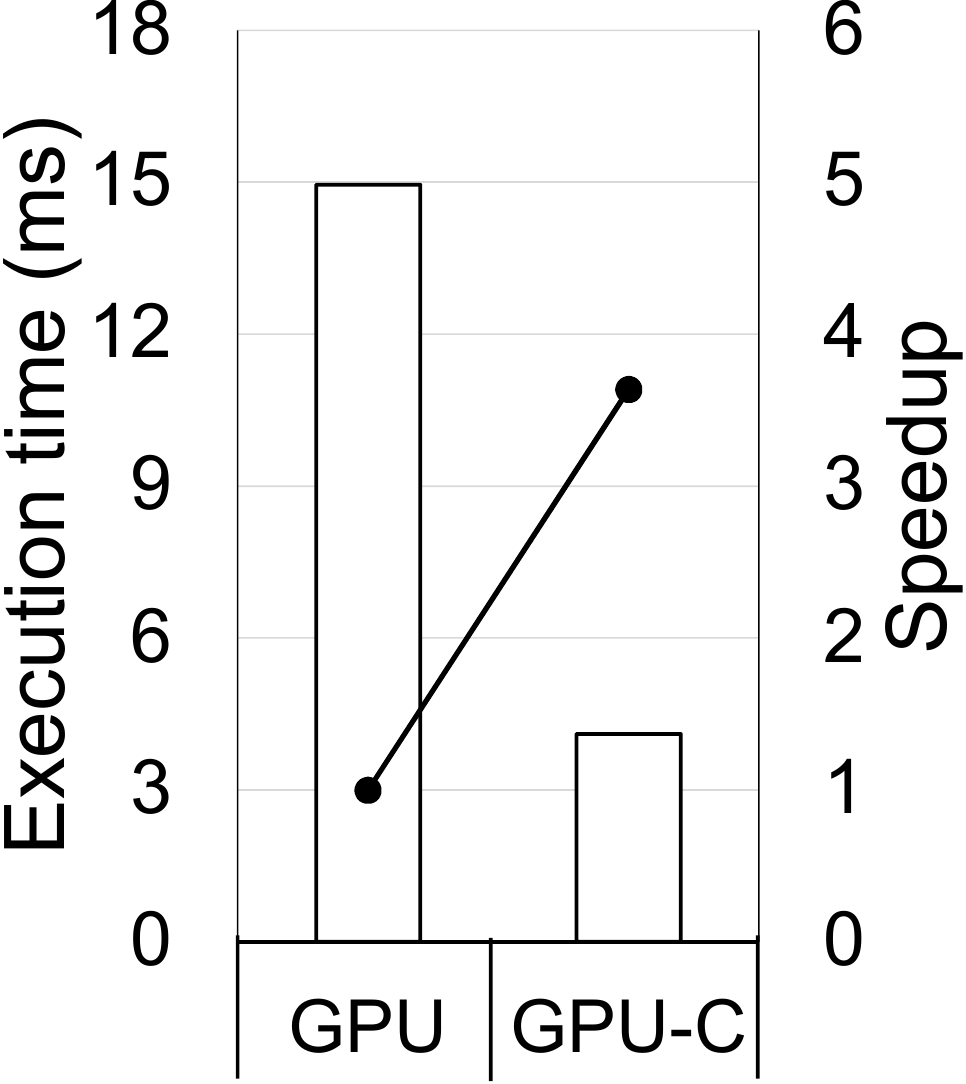}}
  \hspace{0.05in}
  \subfloat[iCRT optimization]{\includegraphics[height=1.3in]{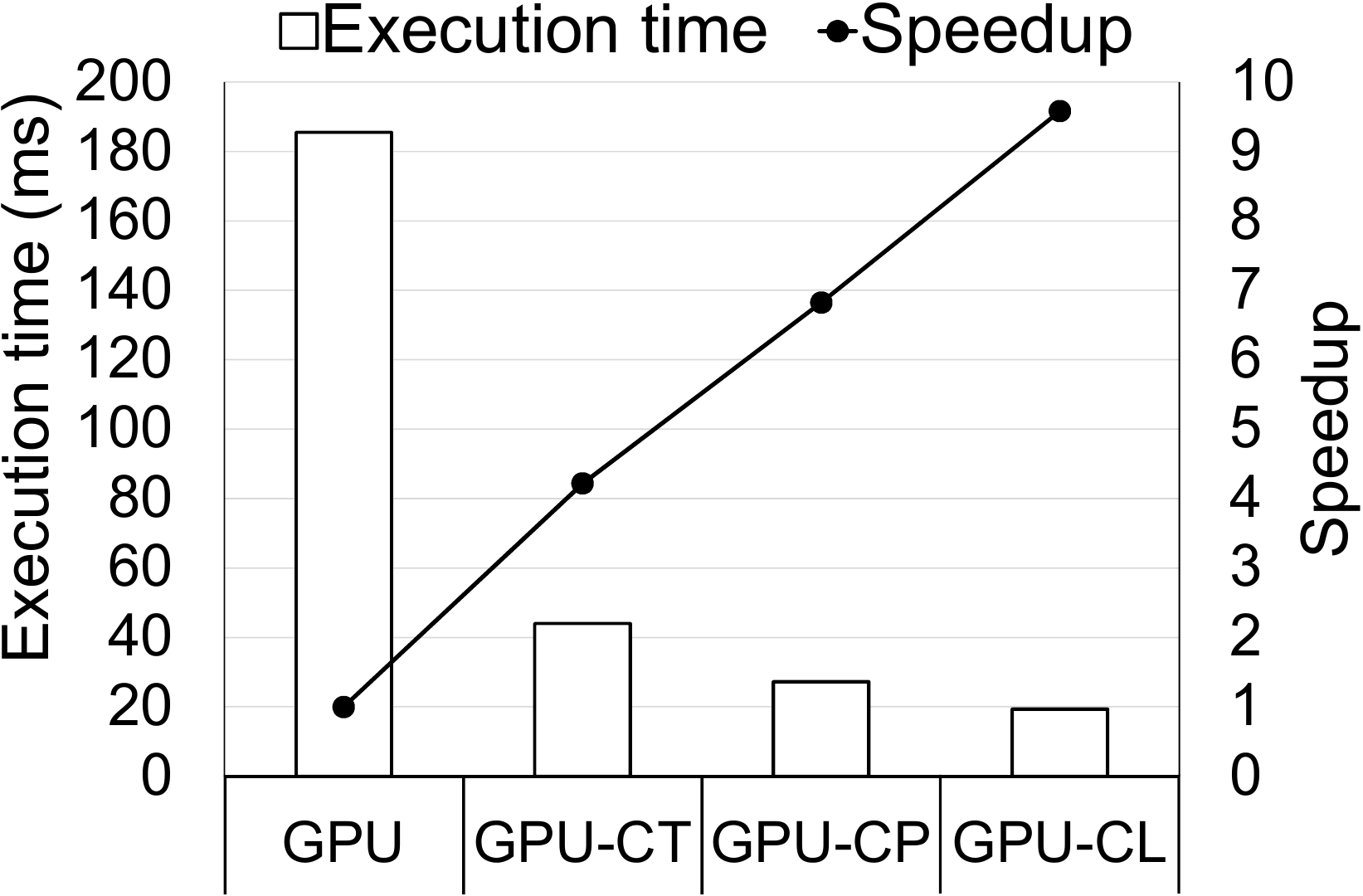}}
  \hspace{0.05in}
  \subfloat[NTT/iNTT optimization]{\includegraphics[height=1.2in]{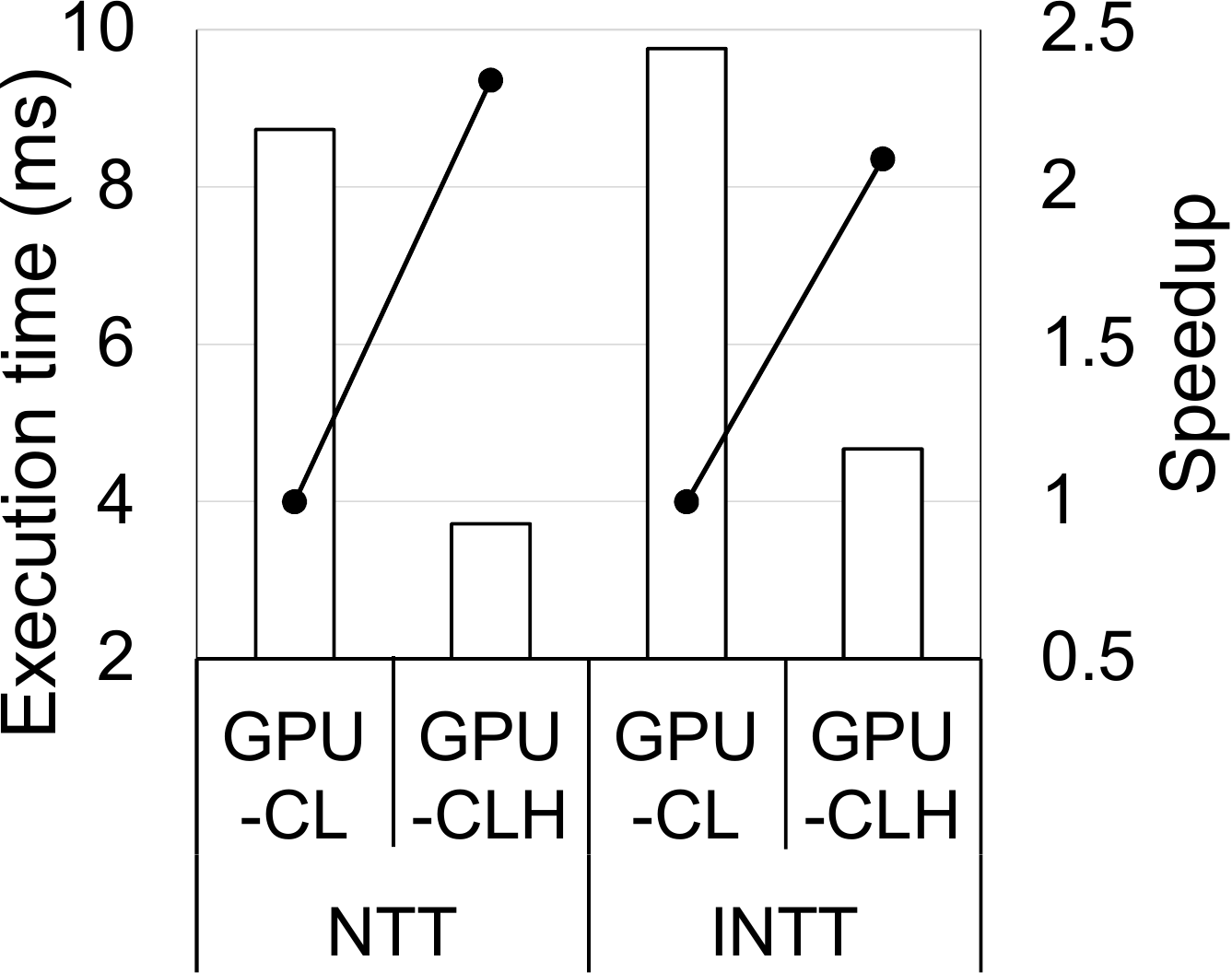}}
  \caption{
    Comparing HE Mul execution time (a) among %the reference HEAAN with 24 threads
    $\textbf{Ref}$-24, the baseline GPU ($\textbf{GPU}$), and
    optimized $\textbf{GPU}$ ($\textbf{GPU-C}$, $\textbf{GPU-CL}$ and $\textbf{GPU-CLH}$),
    and relative speedup of
    (b) $\textbf{GPU}$ and $\textbf{GPU-C}$ for \texttt{CRT},
    (c) $\textbf{GPU}$ and $\textbf{GPU-C[T/P/L]}$ for \texttt{iCRT}, and
    (d) $\textbf{GPU-CL}$ and $\textbf{GPU-CLH}$ for \texttt{NTT/iNTT}.
  }
  \label{fig:eval-gpu}
  \vspace{-0.08in}
\end{figure*}

\emph{Third, the performance of GPU reaches close to full potential through 
our microarchitecture-aware optimizations.}
Figure~\ref{fig:eval-gpu}(a) shows the execution time of HE Mul on various GPU
implementations compared to that of the reference HEAAN running on CPU 
with 24 threads ($\textbf{Ref}$-24).
Even the baseline GPU implementation ($\textbf{GPU}$) outperforms $\textbf{Ref}$-24 
by 1.52$\times$.
Most of the speedup comes from accelerating \texttt{NTT} and \texttt{iNTT}.
\texttt{iCRT} in $\textbf{GPU}$, whose implementation we adopt 
from cuHE~\cite{iccis-2015-cuhe}, performs poorly; it is even 1.43$\times$
slower than that in $\textbf{Ref}$-24 and takes 81.2\% of total HE Mul
execution time.

To reduce the execution time of \texttt{iCRT}, we devised the following
optimizations and compared their performance in Figure~\ref{fig:eval-gpu}(c). 
By adjusting the number of launching threads, we reduced the degree of
performance impact due to cache thrashing, achieving a speedup of
4.22$\times$ ($\textbf{GPU-CT}$) compared to that of $\textbf{GPU}$. 
Pinning thread-local arrays to L1 cache ($\textbf{GPU-CP}$) performs
better than $\textbf{GPU-CT}$, reaching 6.79$\times$ of speedup.
Finally, $\textbf{GPU-CL}$ was the best among all the \texttt{iCRT} 
optimizations (9.58$\times$ of speedup), by effectively exploiting 
$N$$\cdot$$np$-degree parallelism through the loop reordering 
explained in Section~\ref{sec:contribution-3}.

\begin{table}[tb!]
  \centering
  \caption{Comparing the execution time of \texttt{CRT} when
  applying different strategies to RNS conversion.
  $\textbf{GPU-Modx}$ means one modulo operation is applied at every $x$ iteration.}
  \small
  \begin{tabular}{lS[table-number-alignment = center]c}
    \toprule
    & \multicolumn{1}{c}{Time for \textbf{CRT} (ms)} & Speedup \\
    \midrule
    GPU & 14.95 & 1.00$\times$ \\
    GPU-Mod1 & 16.80 & 0.89$\times$ \\
    GPU-Mod2 & 10.47 & 1.43$\times$ \\
    GPU-Mod4 & 7.56 & 1.98$\times$  \\
    GPU-C & 4.11 & 3.64$\times$  \\
    \bottomrule
  \end{tabular}
  \label{tab:eval-crt}
\end{table}

For \texttt{CRT}, performing fewer modulo operations led to a better performance.
Table~\ref{tab:eval-crt} compared the execution time of \texttt{CRT} when using
different strategies to convert a BigInt number to an array of residue numbers 
in an RNS domain. 
$\textbf{GPU}$ does not precompute $\bmod(\beta^k, p_j)$ and performs a modulo
operation on every limb (one limb for $\beta$) of the BigInt.
$\textbf{GPU-Modx}$ means applying modulo operation every $x$ iteration 
in the inner-most loop of line 5 in Algo.~\ref{algorithm:crt}.
Even if using the precomputed table, conducting a modulo operation per
iteration ($\textbf{GPU-Mod1}$) performs even worse than $\textbf{GPU}$ by 1.1$\times$.
This is because both implementations compute the same number of modulo
operations whereas $\textbf{GPU-Mod1}$ imposes more pressure on local memory
utilization.
Fewer modulo operations led to better performance; $\textbf{GPU-Mod4}$ performs
1.98$\times$ better than $\textbf{GPU}$.
By letting the partial sum ($accum$) span three words instead of two words 
utilizing ADC in every iteration,
$\textbf{GPU-C}$ performs best with the speedup of 3.64$\times$. 

\begin{table}[tb!]
  \centering
  \caption{Comparing the execution time of \texttt{NTT} and \texttt{iNTT}
  when altering their radix values.}
  \small
  \begin{tabular}{lcccc}
    \toprule
    & \multicolumn{2}{c}{NTT} & \multicolumn{2}{c}{iNTT} \\
    & Time (ms) & Speedup & Time (ms) & Speedup \\
    \midrule
    Radix-2  & 8.73 & 1.00 & 9.76 & 1.00 \\
    Radix-4  & 4.74 & 1.84 & 5.47 & 1.78 \\
    Radix-16 & 3.65 & 2.39 & 4.88 & 2.00 \\
    Radix-32 & 3.72 & 2.35 & 4.67 & 2.09 \\
    \bottomrule
  \end{tabular}
  \label{tab:radix-exp}
\end{table}

Figure~\ref{fig:eval-gpu}(d) shows the performance improvement of \texttt{NTT}
and \texttt{iNTT} by increasing their radix values.
Because \texttt{NTT} and \texttt{iNTT} algorithms are mostly symmetrical, they
experience a similar degree of performance improvement with 2.35$\times$
(\texttt{NTT}) and 2.09$\times$ (\texttt{iNTT}). 
\texttt{iNTT} improves slightly less because \texttt{iNTT} includes additional 
computations such as element-wise division by $N$, which does not gain speedup
from using higher radix.
Table~\ref{tab:radix-exp} shows the execution time of \texttt{NTT} and 
\texttt{iNTT} when varying the radix values.
As the radix grows from 2 to 16, the performance of \texttt{NTT} and
\texttt{iNTT} increases, taking advantage of reducing the memory accesses and 
hence alleviating the memory bandwidth bottleneck. 
However, the speedup saturates when the radix rises from 16 to 32 because of 
the register spilling to local memory; when radix increases beyond 32, the 
performance deteriorates due to the higher register pressure.

\section{Discussion}
\label{sec:discussion}

\begin{table}[!tb]
  \center
  \caption{The number of required AVX-512 instructions (add, sub, mul, shift, and cmp) for
  each function with ($np$, $qLimbs$) of (43, 19). We compared the cases where each of 64-bit 
  mul, modular mul, and ADC is supported by emulation and
  by a single native instruction.}
  \small
  \vspace{0in}
  \begin{tabular}{lcccc}
    \toprule
    & $\texttt{CRT}$ & $\texttt{NTT}$ & $\texttt{iNTT}$ & $\texttt{iCRT}$ \\
    \midrule
    by emulation & 155M & 48M & 47M & 319M \\
    by single native instr. & 27M & 17M & 17M & 51M \\
    \bottomrule
  \end{tabular}
  \label{tab:emul-instruction}
\end{table}

\noindent
\textbf{The impact of supporting the emulated operations natively on AVX-512:}
As described in Section~\ref{sec:contribution-3}, the cost of emulating 64-bit
mul, modular mul, and ADC operations is significant in
AVX-512 implementations.
If some future CPUs support these instructions natively, the execution time of HE Mul could
be reduced substantially.
Table~\ref{tab:emul-instruction} summarizes the number of AVX-512 instructions
required to perform each major function by comparing the cases where each of 
64-bit mul, modular mul, and ADC is supported by either
emulation and by a single native instruction.
\texttt{CRT} and \texttt{iCRT} require just 17.3\% and 15.8\% of AVX-512 
instructions if a future CPU supports these instructions natively, not 
through emulation.
\texttt{NTT} and \texttt{iNTT} require one third of instructions by the
instruction extension.
Because HE Mul is mostly computation-bound, the significant
reduction in the number of instructions would lead to a similar degree
of performance improvement.

\begin{figure}[!tb]
  \center
  \includegraphics[width=3.4in]{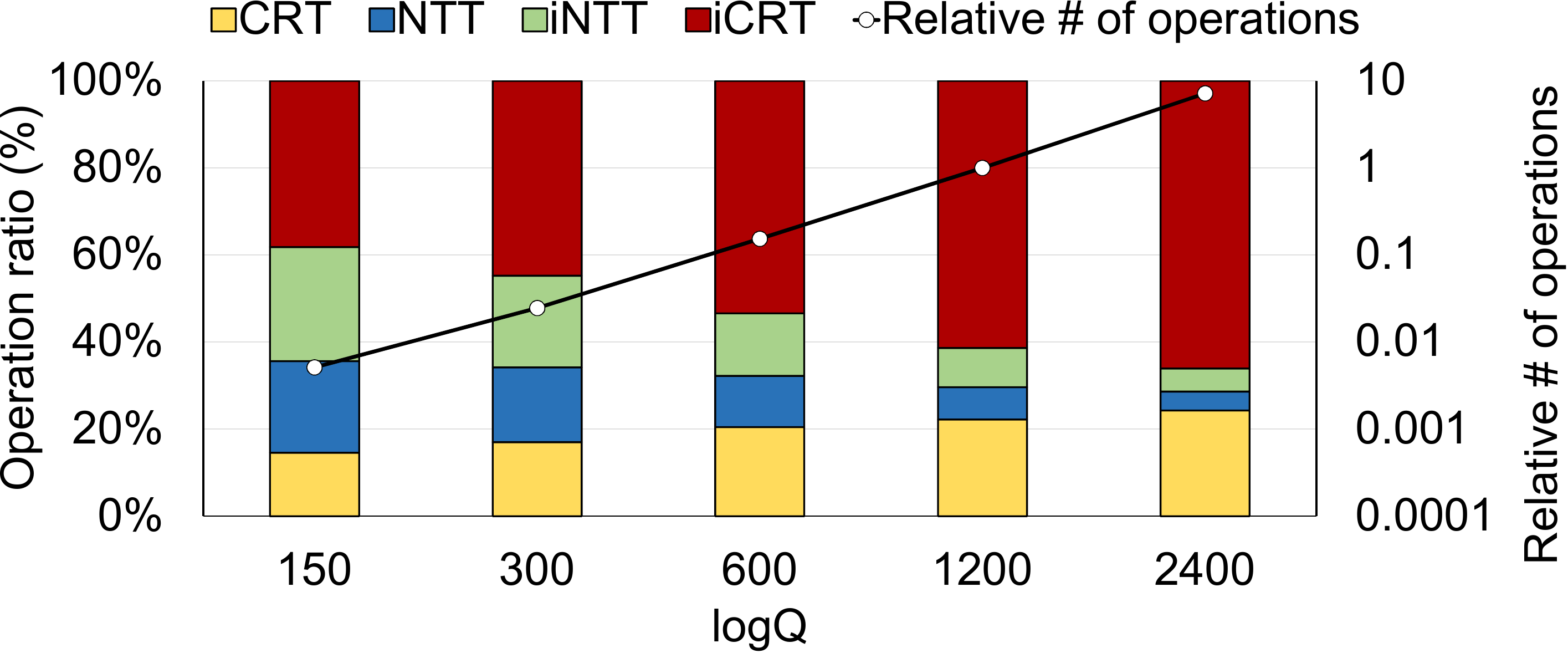}
  \vspace{0in}
  \caption{
    Distribution on the number of operations across functions and the relative number of
    operations for HE Mul on various log$Q$ values.
    The reference log$Q$ is 1,200.
  }
  \label{fig:logQ_op}
  \vspace{-0.1in}
\end{figure}

\noindent
\textbf{Impact of $Q$ on the characteristics of HE Mul:}
$Q$ determines multiplicative depth $L$; a larger depth requires a bigger $Q$.
However, $N$ must increase proportionally to log$Q$ to ensure a certain level
of security (see Table~\ref{tab:Q-N}).
Also, $qLimbs$, $np$, and $PLimbs$ increase in proportion to log$Q$.
Based on these relationships, the computational complexity of 
Table~\ref{tab:num-operations} can be expressed in terms of log$Q$.
The complexity of \texttt{CRT} and \texttt{iCRT} is $\mathcal{O}((\log Q)^3)$
whereas that of \texttt{NTT} and \texttt{iNTT} is 
$\mathcal{O}(\log (\log Q) \cdot (\log Q)^2)$.
Figure~\ref{fig:logQ_op} shows the estimated number of operations for HE
Mul according to log$Q$.
When $Q$ is small (e.g., log$Q$=150), \texttt{CRT}, \texttt{NTT}, \texttt{iNTT},
and \texttt{iCRT} require a similar number of operations, but as $Q$ increases,
\texttt{CRT} and \texttt{iCRT} become more dominant.
Overall, the total number of operations for HE Mul is proportional to 
$(\log Q)^3$.
When an application requires a large number (e.g., billions) of HE Mul, using
large $Q$ amortizes the cost of the expensive bootstrapping operation.
However, using too large $Q$ is costly because the maximum number of messages
($n$) that can be multiplied together by a HE Mul is $\sfrac{N}{2}$, where the
complexity of a HE Mul is super-linear, $\mathcal{O}((\log Q)^3) = \mathcal{O}(N^3)$.
The $Q$ value we mainly target is 1,200, which is large enough to amortize the
cost of bootstrapping; other HE accelerators~\cite{arxiv-2019-heax, ieee-2019-fpga}
focused on much smaller $Q$ values.
For example, \cite{ieee-2019-fpga} used the $Q$ value of 180 without
considering bootstrapping.

\begin{figure}[!tb]
  \center
  \includegraphics[width=3.4in]{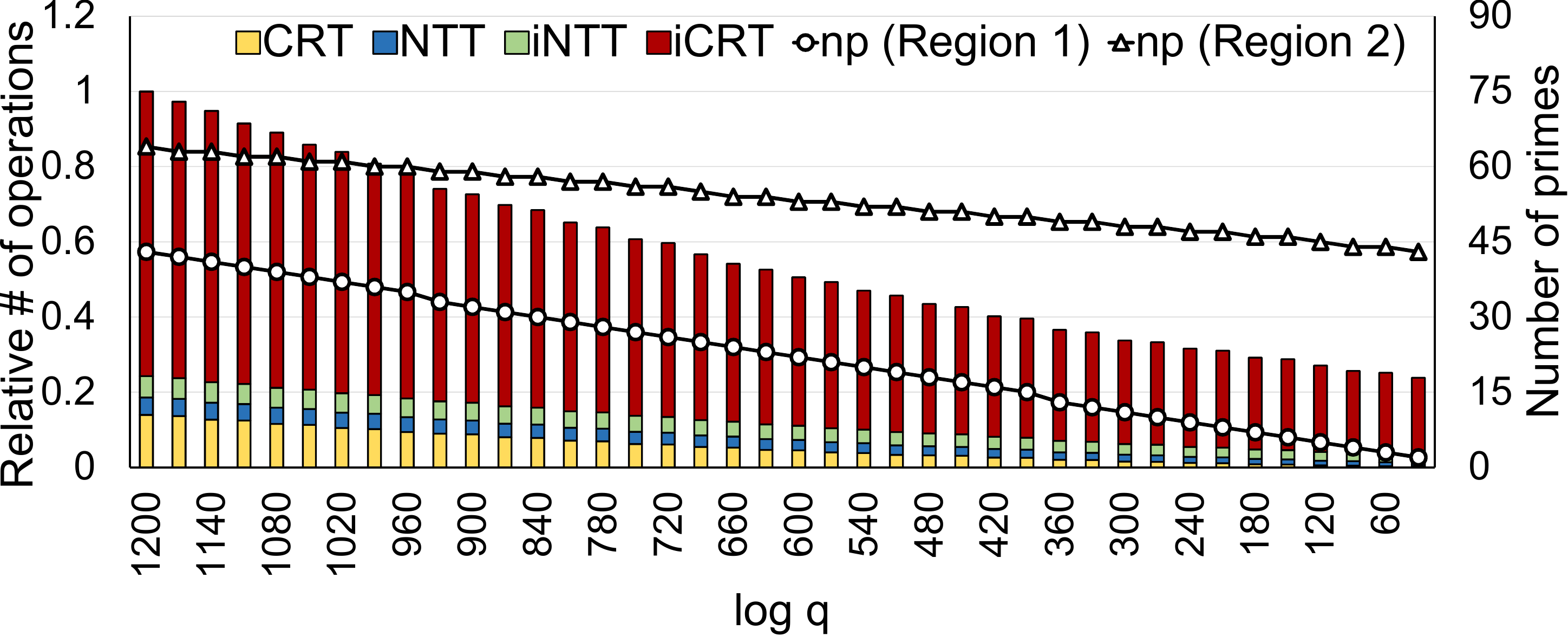}
  \vspace{0in}
  \caption{
    The size of $np$ and the relative number of operations per function according 
    to log$q$.
  }
  \label{fig:logq_op}
  \vspace{0in}
\end{figure}

\noindent
\textbf{Impact of $q$ on the characteristics of HE Mul:}
As described in Section~\ref{sec:introduction-to-heaan}, rescaling, which 
decreases log$q$ by log$p$, is performed after each HE Mul to prevent the
amount of message information in the ciphertext from increasing exponentially.
%
%Therefore,
As HE Mul is repeated, $q$ decreases, so does $qLimbs$ and $np$.
In region 1 of HE Mul, $np$ decreases linearly with log$q$ because
two log$q^2$-bit BitInt numbers are multiplied per ciphertext coefficient.
By contrast, in region 2, $np$ should be set to represent 
(log$q$ + log$Q^2$)-bit BigInt (to multiply over the evaluation key
polynomial).
$PLimbs$ has the same trend as $np$.
%
% Referring to Table~\ref{tab:num-operations}, which describes the amount 
% of computation of each function in terms of $N$, $np$, $qLimbs$, and $PLimbs$, 
% the number of operations required for each function can be estimated according 
% to the change of log$q$.
%
Figure~\ref{fig:logq_op} shows the computation amount of HE Mul 
according to the log$q$ in the AVX-512 configuration of 
Section~\ref{sec:experimental-setup} calculated based on 
Table~\ref{tab:num-operations}.
As $np$ is proportional to (log$q$ + log$Q^2$) in region 2,
the number of operations for HE Mul when log$q$ becomes 30 (the
smallest number where no more HE Mul is applicable) is still
24\% of that when log$q$ is 1,200.
Also, \texttt{iCRT} is dominant regardless the size of $q$.

\begin{table}[tb!]
  \centering
  \caption{Comparing CKKS schemes}
  \small
  \begin{tabular}{lcc}
    \toprule
    \multirow{2}*{CKKS type} & Full-RNS & Non-RNS \\
    & \cite{icfcds-2017-seal,repo-2020-palisade,repo-2020-lattigo} & \cite{ictacis-2017-heaan,repo-2020-helib} \\
    \midrule
    Parameter ($p$ and $Q$) selection  & Rigid & Flexible \\
    Multiplicative depth & Lower & Deeper \\
    HE operation speed & Faster & Slower \\
%    Approximate error & Larger & Smaller \\
    \multirow{2}*{Main function of HE Mul} & \texttt{NTT}, \texttt{iNTT}, & \texttt{CRT}, \texttt{iCRT}, \\
    & mod up/down & \texttt{NTT}, \texttt{iNTT} \\
    \bottomrule
  \end{tabular}
  \label{tab:comp-ckks}
\end{table}

\noindent
\textbf{The applicability of optimization techniques to other libraries supporting CKKS:}
In addition to HEAAN, there are several other
libraries~\cite{repo-2020-helib, icfcds-2017-seal, repo-2020-palisade, repo-2020-lattigo} that
support CKKS.
The libraries fall into one of the two groups: libraries supporting full-RNS and non-RNS CKKS,
as shown in Table~\ref{tab:comp-ckks}.
Non-RNS types manage each coefficient of polynomials of ciphertext in a big integer domain.
By contrast, full-RNS types let the coefficient stay in an RNS domain during the whole
procedure of homomorphic operations, resulting in faster operations.
However, using full-RNS is less flexible because there exist rigid limitations while
choosing the rescaling factor $p$ and the modulus $Q$.
Each prime composing an RNS
representation of $Q$ in full-RNS CKKS should be set to be close to the rescaling factor
$p$~\cite{icsac-2018-rnsheaan}.
In non-RNS CKKS, by contrast, one can freely choose a rescaling factor $p$ independent of $Q$
without considering the approximation error existing in full-RNS CKKS.
Moreover, due to these parameter limitations, the multiplicative depth of full-RNS variants is
lower than an non-RNS scheme for a given security bit and error bound.

For the libraries supporting CKKS other than HEAAN, our optimizations described in
Section~\ref{sec:contribution-3} are partially applicable.
Other non-RNS CKKS libraries can apply our optimization techniques because they use the
same primary functions as HEAAN.
Also, although full-RNS variants do not require \texttt{CRT} and \texttt{iCRT},
the optimization techniques for \texttt{NTT} and
\texttt{iNTT} can be applied.
Also, we can partially apply our techniques in CRT to the functions that change the
number of primes in the RNS domain (\texttt{mod up/down} in Table~\ref{tab:comp-ckks}).
\texttt{mod up} increases the number of primes of an RNS representation of a given big integer,
whereas \texttt{mod down} decreases it with an additional division
operation~\cite{icsac-2018-rnsheaan, icsac-2016-rnsbfv}.
Both functions can take advantage of our optimizations as their core functions are similar
to applying \texttt{CRT} right after \texttt{iCRT}.

\section{Related Work}
\label{sec:relatedwork}

\textbf{FPGA-based HE accelerators:}
There have been numerous studies~\cite{iwches-2015-ltv,ieee-2016-customaccelerator,ieee-2016-coprocessor,ieee-2018-hepcloud,ieee-2019-fpga,arxiv-2019-heax} 
to accelerate HE operations using FPGA. 
\cite{iwches-2015-ltv,ieee-2016-customaccelerator,ieee-2016-coprocessor} 
accelerate LTV-based FHE schemes whereas
\cite{ieee-2018-hepcloud,ieee-2019-fpga} acclerate FV-based FHE schemes.
However, LTV and FV schemes have a limitation in practical use because they 
cannot perform approximate computations.
HEAX~\cite{arxiv-2019-heax} uses FPGA to accelerate Microsoft SEAL, which 
supports a full-RNS variant of the CKKS scheme; 
however, HEAX considers only small parameter sizes ($Q\leq 438$ and $N\leq 2^{14}$),
and the full-RNS variant it targets is not as versatile as the original HEAAN
we accelerate in this paper due to the limitations in choosing rescaling
factors and prime numbers.

\textbf{GPU libraries for HE:}
\cite{iccis-2015-cuhe,gpgpu-2016-yashe,iacr-2018-fvgpu,ieee-2019-rnsvariants} 
propose to accelerate the HE operations using GPUs.
However, they either do not take advantage of the algorithm's internal 
parallelism sufficiently, operate on only small or limited parameters,
or do not consider the cost of modulo operations in GPUs.
Moreover, all the aforementioned studies 
did not conduct a rigorous analysis of computational
complexity and data access patterns of HE Mul, making it hard to assess the
effectiveness of the proposed accelerators compared to the CPU or GPU
implementations applying architecture-aware optimizations.

\section{Conclusion}
\label{sec:conclusion}

We have demystified the key operations of HEAAN, a representative and popular 
FHE scheme, from a computer architect's perspective.
After identifying that multiplying the encrypted data (ciphertext) is the most 
computationally demanding, we accelerated the major functions of HE
Mul (\texttt{CRT}, \texttt{NTT}, \texttt{iNTT}, and \texttt{iCRT}) 
on CPU and GPU. 
To accelerate the major functions on CPU, we populate multiple cores by using 
multi-threading (inter-core parallelism) and AVX-512 instructions 
(intra-core parallelism).
We accelerate HE Mul on GPU by effectively exploiting massive thread-level 
parallelism. 
Moreover,
based on the in-depth analysis of the major functions for HE Mul, 
we introduced a series of architecture-aware optimization techniques 
such as loop reordering and matrix transposition for \texttt{iCRT}
and taking a synergy between precomputation and delayed modulo
operations for \texttt{CRT}.
Our accelerated HEAAN on CPU and GPU outperforms the reference single-thread
HEAAN on CPU by 42.9$\times$ and 134.1$\times$, respectively, in HE Mul,
setting a new baseline for HE acceleration studies targeting practical usage.
%

%%%%%%% -- PAPER CONTENT ENDS -- %%%%%%%%

%%%%%%%%% -- BIB STYLE AND FILE -- %%%%%%%%
\bibliographystyle{IEEEtranS}
\bibliography{ref}
%%%%%%%%%%%%%%%%%%%%%%%%%%%%%%%%%%%%

\end{document}